\documentclass[aps,pra,twocolumn,superscriptaddress,longbibliography,nobalancelastpage]{revtex4-2}

\usepackage{graphicx}
\usepackage{hyperref}
\usepackage{amsmath}
\usepackage{amssymb}
\usepackage{tikz}
\usepackage{float}	
\usepackage{tabularx}  
\usepackage{booktabs}   
\usepackage{tikz}
\usepackage[caption=false]{subfig}
\usepackage[percent]{overpic}
\newcommand{\ket}[1]{\left|#1\right\rangle}

\begin{document}

\title{
Phase-Space Topology in a Single-Atom Synthetic Dimension
}

\author{Kyungmin Lee}
\affiliation{\mbox{Department of Computer Science and Engineering, Seoul National University, Seoul 08826, Republic of Korea}}
\affiliation{\mbox{Automation and System Research Institute, Seoul National University, Seoul 08826, Republic of Korea}}
\affiliation{\mbox{NextQuantum, Seoul National University, Seoul 08826, Republic of Korea}}

\author{Sunkyu Yu}
\affiliation{Intelligent Wave Systems Laboratory, Department of Electrical and Computer Engineering, Seoul National University, Seoul 08826, Republic of Korea}

\author{Jiyong Kang}
\affiliation{\mbox{Department of Computer Science and Engineering, Seoul National University, Seoul 08826, Republic of Korea}}
\affiliation{\mbox{Automation and System Research Institute, Seoul National University, Seoul 08826, Republic of Korea}}
\affiliation{\mbox{NextQuantum, Seoul National University, Seoul 08826, Republic of Korea}}

\author{Seungwoo Yu}
\affiliation{\mbox{Department of Computer Science and Engineering, Seoul National University, Seoul 08826, Republic of Korea}}
\affiliation{\mbox{Automation and System Research Institute, Seoul National University, Seoul 08826, Republic of Korea}}
\affiliation{\mbox{NextQuantum, Seoul National University, Seoul 08826, Republic of Korea}}

\author{Wonhyeong Choi}
\affiliation{\mbox{Department of Computer Science and Engineering, Seoul National University, Seoul 08826, Republic of Korea}}
\affiliation{\mbox{Automation and System Research Institute, Seoul National University, Seoul 08826, Republic of Korea}}
\affiliation{\mbox{NextQuantum, Seoul National University, Seoul 08826, Republic of Korea}}

\author{Daun Chung}
\affiliation{\mbox{Department of Computer Science and Engineering, Seoul National University, Seoul 08826, Republic of Korea}}
\affiliation{\mbox{Automation and System Research Institute, Seoul National University, Seoul 08826, Republic of Korea}}
\affiliation{\mbox{NextQuantum, Seoul National University, Seoul 08826, Republic of Korea}}

\author{Sumin Park}
\affiliation{\mbox{Department of Computer Science and Engineering, Seoul National University, Seoul 08826, Republic of Korea}}
\affiliation{\mbox{Automation and System Research Institute, Seoul National University, Seoul 08826, Republic of Korea}}
\affiliation{\mbox{NextQuantum, Seoul National University, Seoul 08826, Republic of Korea}}

\author{Taehyun Kim}
\email[]{taehyun@snu.ac.kr}
\affiliation{\mbox{Department of Computer Science and Engineering, Seoul National University, Seoul 08826, Republic of Korea}}
\affiliation{\mbox{Automation and System Research Institute, Seoul National University, Seoul 08826, Republic of Korea}}
\affiliation{\mbox{NextQuantum, Seoul National University, Seoul 08826, Republic of Korea}}
\affiliation{\mbox{Institute of Applied Physics, Seoul National University, Seoul 08826, Republic of Korea}}

\begin{abstract}
We investigate topological features in the synthetic Fock-state lattice (FSL) of a single-atom system described by the quantum Rabi model. 
By diagonalizing the Hamiltonian, we identify a zero-energy defect state localized at a domain wall of the FSL, whose spin polarization is topologically protected.
To address the challenge of applying band topology to the FSL, we introduce a physically motivated and directly measurable topological invariant based on phase-space geometry---the phase-space winding number.
We show that the Zak phase, computed using a phase-space parameter, is related to the invariant. 
This quantized geometric phase reflects the spin polarization of the defect state, demonstrating a bulk--boundary correspondence. 
The resulting phase-space topology reveals the emergence of single-atom dressed states with contrasting properties---topologically protected spin states and driving-tunable bosonic states. 
Our results establish phase-space topology as a novel framework for exploring topological physics in single-atom synthetic dimensions, uncovering quantum-unique topological protection distinct from classical analogs.
\end{abstract}

\maketitle

\textit{Introduction}---A single two-level atom interacting with a harmonic oscillator, as described by the quantum Rabi model (QRM) \cite{RabiModel} or the Jaynes--Cummings (JC) model \cite{JCmodel}, has been a prominent subject of theoretical and experimental studies \cite{larson_jaynescummings_2024,QRM.Review}. 
Along with the quasi-exact solvability of the QRM \cite{Braak.PRL.107.100401,quasi_QRM}, various theoretical \cite{QRM_phasetransition} and experimental \cite{QRM_iontrap, QRM_nmr} studies have explored the quantum phase transition in the QRM. 
Subsequent studies have investigated variants of the QRM incorporating anisotropy \cite{QRM_anisotropic}, dissipation \cite{QRM_dissipative}, nonlinearity \cite{QRM_nonlinear}, and finite-frequency effects \cite{ying_quantum_2022}, each revealing distinct physical phenomena. 
In parallel, recent experimental advances have enabled the realization of ultra-strong \cite{circuit_usc, circuit_usc2, circuit_usc3} and deep-strong coupling regimes \cite{dsc_photonic,dsc_circuit,dsc_ion,dsc_ion2, dsc_ion3} in the QRM \cite{QRM.Review}, allowing the simulation of exotic phenomena emerging from strong atom–oscillator coupling.

Since the discovery of the quantum Hall effect \cite{quantumHall}, topological phases of matter have been extensively studied across a variety of physical platforms, including optical lattices \cite{topo_lattice, topo_lattice2}, photonics \cite{topo_photonics, topo_photonics2}, trapped ions \cite{topo_ion, topo_ion2}, superconducting circuits \cite{topo_superconducting, topo_superconducting2}, and neutral atoms \cite{topo_neutral, topo_neutral2}. 
An emerging topic is to realize higher-dimensional topological phenomena within physically accessible platforms, for example, using the concept of synthetic dimensions \cite{synthetic_photon, synthetic_photon2, synthetic_ion, synthetic_ion2, fsl_photon}. 
While most studies of synthetic-dimensional physics have focused on classical platforms, using the Fock basis in quantum systems to form Fock-state lattices (FSLs) has recently been investigated \cite{fsl,fsl2,fsl3,fsl4,fsl5}.

In the absence of spin coupling, the QRM can be viewed as a semi-infinite one-dimensional (1D) FSL \cite{fsl2}, and a relation to an infinite Su--Schrieffer--Heeger (SSH) chain has been suggested \cite{ssh,fsl2}.
While finite 1D FSLs, such as two-mode JC \cite{fsl} and central-spin models \cite{fsl2}, support SSH-like edge modes, the topological properties of the semi-infinite 1D FSL described by the QRM remain unexplored.

Here, we demonstrate topological phenomena in a generalized QRM with spin coupling.  
We derive analytic solutions and identify a zero-energy defect state localized at a domain wall, whose spin polarization is topologically protected.  
To address the absence of translational symmetry in FSLs, we propose a phase-space approach that enables analytic estimation of topological features.  
We define a physically motivated topological invariant---the phase-space winding number---that captures the topological protection of spin polarization.  
We demonstrate that the spin polarization of the defect state is robust to parameter noise, enabling high-fidelity nonclassical bosonic-state generation and seemingly contradictory active quantum functionalities---noise-robust but target-selective spin control with dissipation-induced fidelity enhancement.
We further show that the invariant corresponds to the quantized geometric-phase difference between the two sublattices, reflecting bulk--boundary correspondence.  
Our results provide a novel framework for exploring topological features in a single-atom system, with potential realizations in trapped ions~\cite{dsc_ion, ion_sdf} and superconducting circuits~\cite{sideband_superconducting, sideband_superconducting2, sideband_superconducting3, sideband_superconducting4}.

\begin{figure}[t!]
\centerline{\includegraphics[width=1.\columnwidth]{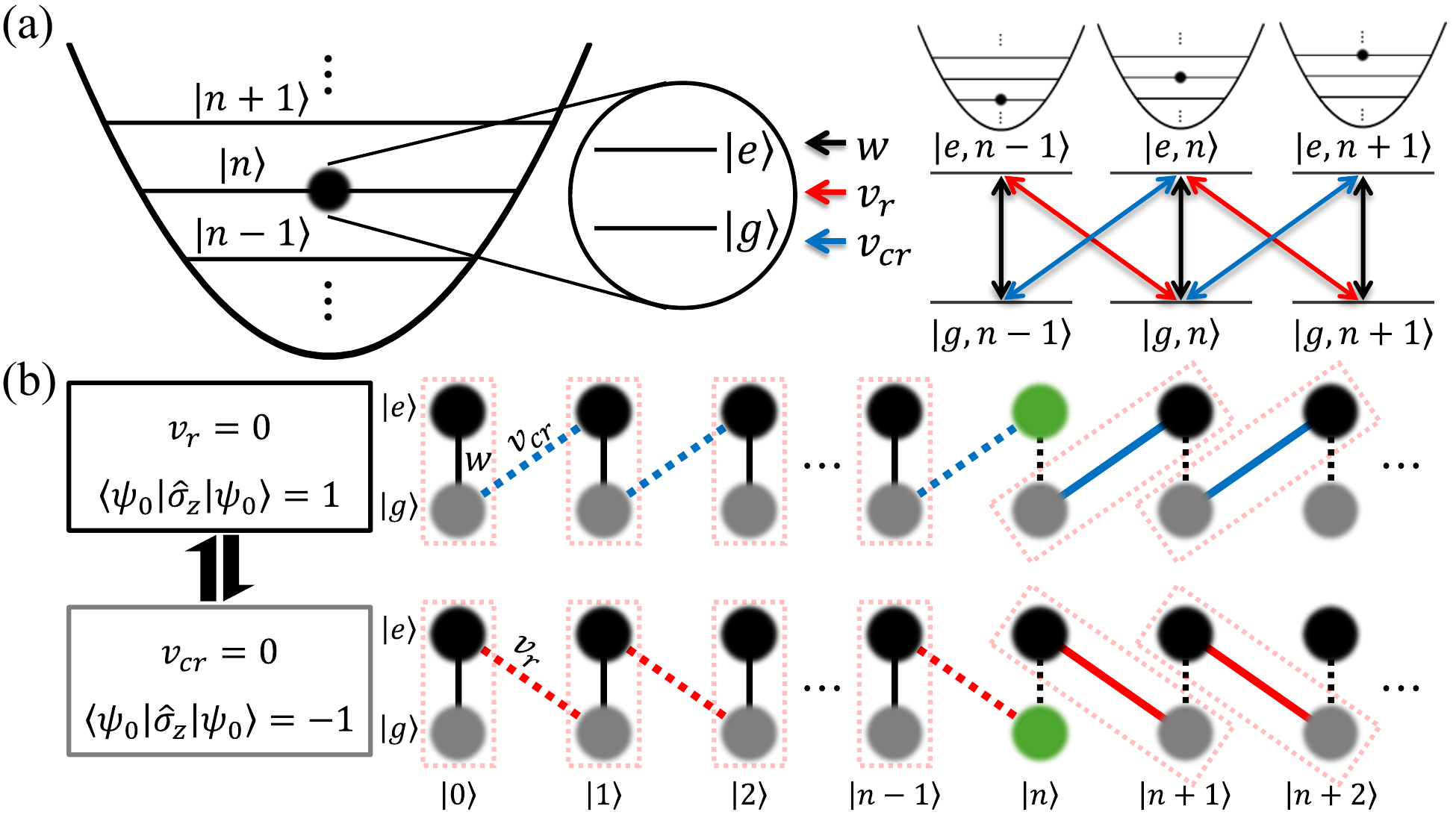}}
\caption{
Illustration of (a) the setup under consideration and (b) a domain wall formed for \(w \neq 0\). 
(b) The upper and lower lattices correspond to \(v_r=0\) and \(v_{cr}=0\), respectively. 
Each site is labeled by a Fock state \(|n\rangle\), with black and gray circles representing the spin states \(|e\rangle\) and \(|g\rangle\). 
Solid and dashed lines indicate the stronger and weaker of the intra-cell (\(w\)) and inter-cell (\(v_r\) or \(v_{cr}\)) couplings, respectively. 
Red dotted boxes mark the dimerized unit cells. 
A domain wall appears at \(|n\rangle\) when \(|v|\sqrt{n}<w<|v|\sqrt{n+1}\), highlighted by a green circle.
}
\label{fig:figure1}
\end{figure}

\textit{Defect states and energy spectrum}---We consider a system described by the generalized quantum Rabi Hamiltonian \cite{general_QRM}, neglecting on-site energy terms; see Fig.~\ref{fig:figure1}(a).
\begin{align}
\hat{H} = w \hat{\sigma}_x + \left( v_r \hat{a}^\dagger \hat{\sigma}_- + v_{cr} \hat{a} \hat{\sigma}_- + \text{h.c.} \right),
\label{eqn:eqn1}
\end{align}
where h.c.\ denotes the Hermitian conjugate. 
\(\hat{\sigma}_x\) is the Pauli \(x\)-operator, \(\hat{a}^\dagger\) (\(\hat{a}\)) is a bosonic creation (annihilation) operator, and operators \(\hat{\sigma}_\pm\) denote the atomic raising and lowering operators acting on a two-level system. 
We treat both \(v_r = |v_r| e^{i\phi_r}\) and \(v_{cr} = |v_{cr}| e^{i\phi_{cr}}\) as complex parameters, while \(w\) is assumed to be real and positive. 
The amplitudes \( |v_r| \) and \( |v_{cr}| \) represent the coupling strengths between the atom and the red- and blue-sideband transitions, respectively, with the phases \( \phi_r \) and \( \phi_{cr} \) denoting their relative phase with respect to the carrier drive \( w \)~\cite{dsc_ion, sideband_superconducting, sideband_superconducting2}.
The Hamiltonian possesses chiral symmetry with respect to the Pauli \(z\)-operator, \(\hat{\sigma}_z\), and can be implemented in trapped-ion~\cite{dsc_ion, ion_sdf} and superconducting-circuit~\cite{sideband_superconducting, sideband_superconducting2, sideband_superconducting3, sideband_superconducting4} platforms.

The Hamiltonian in Eq.~(\ref{eqn:eqn1}) effectively describes an SSH-like, semi-infinite chain in a FSL when either \( v_r = 0 \) or \( v_{cr} = 0 \) (Fig. \ref{fig:figure1}). 
The Fock states \( |n\rangle \) serve as lattice sites, with the hopping amplitude between neighboring sites scaling as \( \sqrt{n} \), lacking the discrete translational symmetry. 
Since \( w \) remains constant across all lattice sites, the increasing hopping amplitude induces a domain wall at the site \( |n\rangle \) where the relative strength between \( w \) and \( |v| \sqrt{n} \) is reversed. Here, \( v \) denotes the non-zero hopping parameter.
In this semi-infinite SSH configuration with broken translational symmetry, the domain wall results in a defect at \( |n\rangle \), as illustrated in Fig.~\ref{fig:figure1}. 

To unveil the dynamics in this synthetic FSL, we derive the analytical solutions of its states and spectrum for the entire regime between two extremes illustrated in Fig. \ref{fig:figure1}.
For a given \(w\), the dynamics over the lattice, including the emergence of defect states, is governed by the one with larger amplitude between \( v_r \) and \( v_{cr} \), while the smaller one acts as a long-range hopping correction to the system \cite{SSH_long_range_hopping}.
By diagonalizing the Hamiltonian in Eq.~(\ref{eqn:eqn1}), we obtain the analytical solution for the complete set of energy eigenstates and eigenvalues for the general case \( |v_r| \neq |v_{cr}| \)~\cite{supplement}:
\begin{align}
|\psi_0\rangle &= \hat{U} |0\rangle |A\rangle, \quad 
|\psi_{n+1}^{\pm}\rangle = \hat{U} \left( \frac{|n{+}1\rangle |A\rangle \pm |n\rangle |B\rangle}{\sqrt{2}} \right), \nonumber \\
E_n^{\pm} &= \pm \sqrt{ \left| \left( | v_r |^2 - |v_{cr} |^2 \right) n \right| },
\label{eqn:eqn2}
\end{align}
for \( n \in \{0, 1, 2, \dots\} \) and \(\hat{U}=\hat{R}(\theta) \hat{D}(\alpha) \hat{S}(r)\). Here, \(|A\rangle = |g\rangle\) for \(|v_r| > |v_{cr}|\), \(|A\rangle = |e\rangle\) for \(|v_r| < |v_{cr}|\), and \(|B\rangle = \hat{\sigma}_x|A\rangle\).
The operators \(\hat{D}(\alpha)\), \( \hat{S}(\xi) \), and \( \hat{R}(\theta) \) are the displacement, squeezing, and phase-shifting operators, respectively~\cite{quantum_optics, phase_shift_operator}, where
\begin{align}
&\alpha = \frac{w (v_{cr}^* - v_r)}{ |v_r|^2 - |v_{cr}|^2 }, \quad \tanh(2r) = \frac{2 |v_r v_{cr}|}{ |v_r|^2 + |v_{cr}|^2 },
\label{eqn:eqn3}
\end{align}
for \(r\in\mathbb{R}\), \( \theta =(\phi_{cr}-\phi_r)/2\), \(\phi=(\phi_r + \phi_{cr})/2\), where \((\cdot)^*\) denotes complex conjugation (see Appendix A for the comparison between analytic expression and numerical results). As shown in Eq.~(\ref{eqn:eqn2}), the weaker of the two couplings (\(v_r\), \(v_{cr}\)) modifies the phase-space structure of the eigenstates by introducing additional displacement, rotation, and squeezing. Moreover, the zero-energy defect state \( |\psi_0\rangle \) maintains a fixed spin polarization, \( \langle \psi_0 | \hat{\sigma}_z | \psi_0 \rangle \).

When \( |v_r| = |v_{cr}| = |v| \), the energy spectrum is:
\begin{align}
E^{\pm}(x, \phi) = \pm \sqrt{ w^2\left( 1 - \cos{2\phi} \right)/2 + 2 |v|^2 x^2 },
\label{eqn:eqn4}
\end{align}
where \( x \) is the eigenvalue of the position operator \( \hat{x}= \frac{1}{\sqrt{2}}\left(\hat{a}+\hat{a}^{\dagger}\right) \)~\cite{supplement}. 
The energy gap closes at \( \phi = m\pi \) for \( m \in \mathbb{Z} \), forming a periodic Dirac-cone structure in a \((1\!+\!1)\)-dimensional space spanned by the phase-space coordinate \( x \) and the parameter \( \phi \) (see Appendix).

\textit{Phase-space topology}---To investigate topological features of our semi-infinite and aperiodic system, indicated by the topologically protected spin polarization of a defect state around a periodic Dirac cone, we develop a theoretical framework of phase-space topology.
This framework is related to prior approaches in dynamical systems theory \cite{bifurcationtheory}, classical optics \cite{phase_optics}, and continuous systems \cite{phase_quant, phase_quant2}.

The spin polarization of the zero-energy defect state is topologically protected and robust against noise in the hopping parameters \(w\), \(v_r\), and \(v_{cr}\). It depends solely on the relative magnitude between \(|v_r|\) and \(|v_{cr}|\): 
\begin{align}
\left\langle \psi_0 \left| \hat{\sigma}_z \right| \psi_0 \right\rangle = \text{sgn}(|v_{cr}| - |v_r|),
\label{eqn:eqn5}
\end{align}
as derived from Eq. (\ref{eqn:eqn2}). Thus, the two regimes \(|v_r| > |v_{cr}|\) and \(|v_r| < |v_{cr}|\) correspond to distinct topological phases, characterized by opposite spin polarizations of the defect state. Since the defect is localized at the interface between two distinct domains (see Fig. \ref{fig:figure1}), it can be interpreted as a boundary property of the lattice.

As Eq.~(\ref{eqn:eqn2}) shows, the difference in energy \(E_n^{\pm}\) and \(E_{n\pm1}^{\pm}\) vanishes in the limit \( n \to \infty \), defining the bulk regime of the lattice. 
In this limit, a conventional momentum-space analysis can be applied approximately, yielding a winding number in analogy with the SSH model, but it provides neither a clear physical interpretation nor an experimentally measurable quantity.
Alternatively, one can apply real-space topological invariants~\cite{realspace_winding,topological_marker}, developed for systems without translational symmetry, to FSLs.
However, these invariants may require local measurements of Fock states \(|n\rangle\) at \(n\) beyond a parameter-dependent cutoff~\cite{supplement}, which is impractical on current quantum hardware, in contrast to real-space systems.

\begin{figure}[t!]
\centerline{\includegraphics[width=1.\columnwidth]{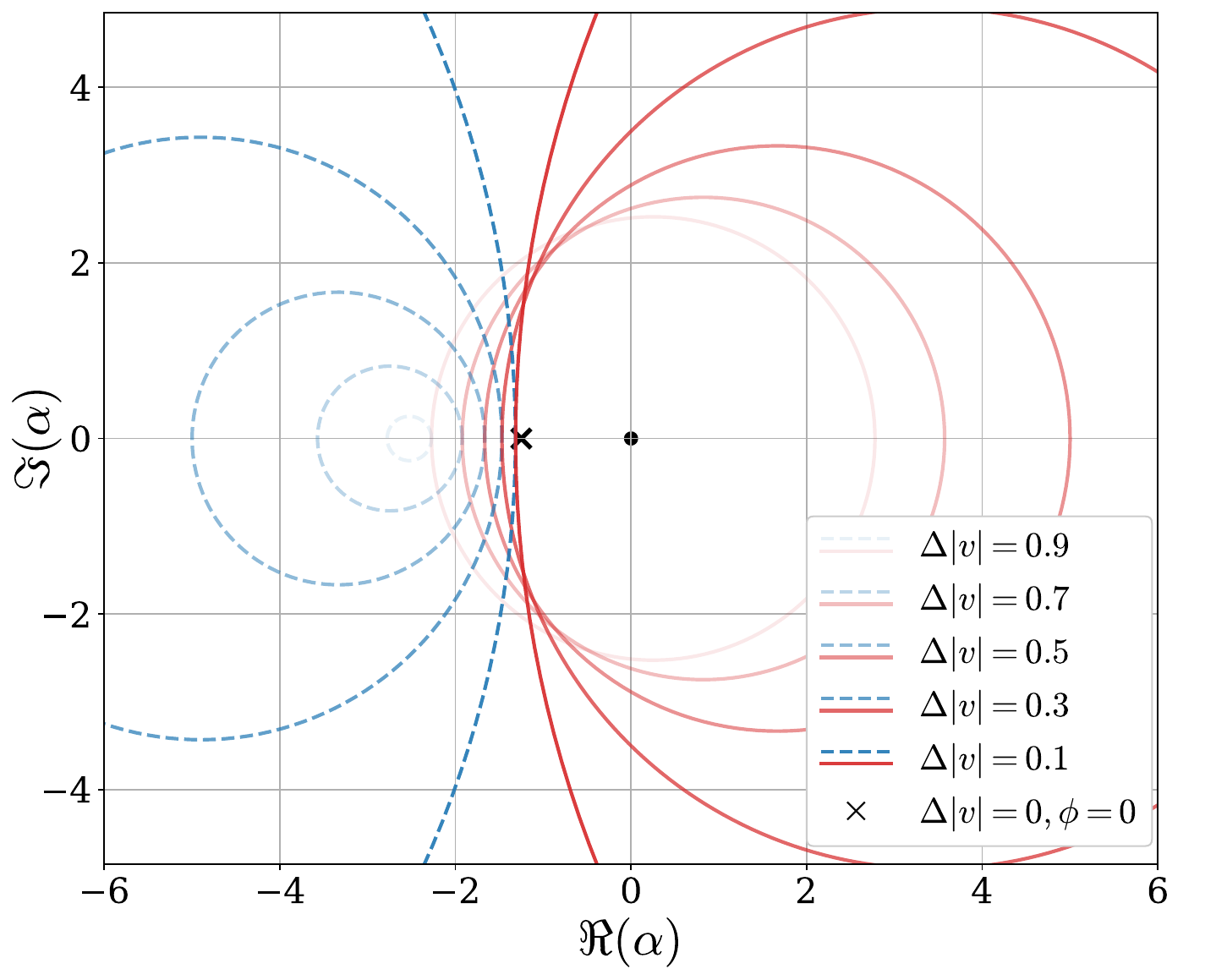}}
\caption{
Phase-space trajectories for \(\Delta |v| \equiv \bigl||v_r|-|v_{cr}|\bigr|\) at \(w=2.5\). 
Red solid (blue dashed) curves correspond to the regime \(|v_r|>|v_{cr}|\) (\(|v_r|<|v_{cr}|\)). 
As \(\Delta |v|\to 0\), both the center and the radius of the trajectories diverge, while the intersection of the two regimes converges to the location marked by a cross. 
The phase-space winding number indicates whether the trajectory encloses the origin.
}
\label{fig:figure2}
\end{figure}

To resolve the difficulty in directly applying band topology to FSLs that lack translational symmetry, we focus on the phase parameter \( \phi \), which provides an alternative periodicity over the phase space.
To remove the additional degree of freedom, we fix \(\phi_{cr}\) and move to a frame rotated by \(e^{i\phi_{cr}}\), such that the eigenstates are centered along a circular trajectory \(\alpha(\phi) = C + R e^{2i\phi}\) in phase space.
As \(\phi\) varies from 0 to \(\pi\), \(\alpha(\phi)\) traces a circular loop in Fig.~\ref{fig:figure2}, corresponding to a helical trajectory of the eigenstates along the \(\theta\)-axis (see Appendix B).
Accordingly, the regimes \( |v_r| > |v_{cr}| \) and \( |v_r| < |v_{cr}| \) correspond to the geometric conditions \( R > |C| \) and \( R < |C| \), respectively.
In analogy with the winding number in the SSH model~\cite{ssh_lecture}, we define a phase-space winding number for the SSH-like chain in the FSL as:
\begin{align}
W = \frac{1}{2\pi i} \int_{0}^{\pi} d\phi \, \frac{d}{d\phi} \ln \alpha(\phi) = 
\frac{1-\text{sgn}(|v_{cr}| - |v_r|)}{2}.
\label{eqn:eqn6}
\end{align}
This quantity indicates whether the trajectory \(\alpha(\phi)\) encloses the origin in phase space, where \( \phi \) substitutes the role of the Brillouin zone in defining the Zak phase in 1D periodic systems. 
The resulting winding number exhibits a discrete jump at the critical point \( |v_r| = |v_{cr}| \), as shown in Fig.~\ref{fig:figure2}. 
See Appendix B for a detailed calculation of \(W\).

\begin{figure}[b!]
\centering
\includegraphics[width=1.\columnwidth]{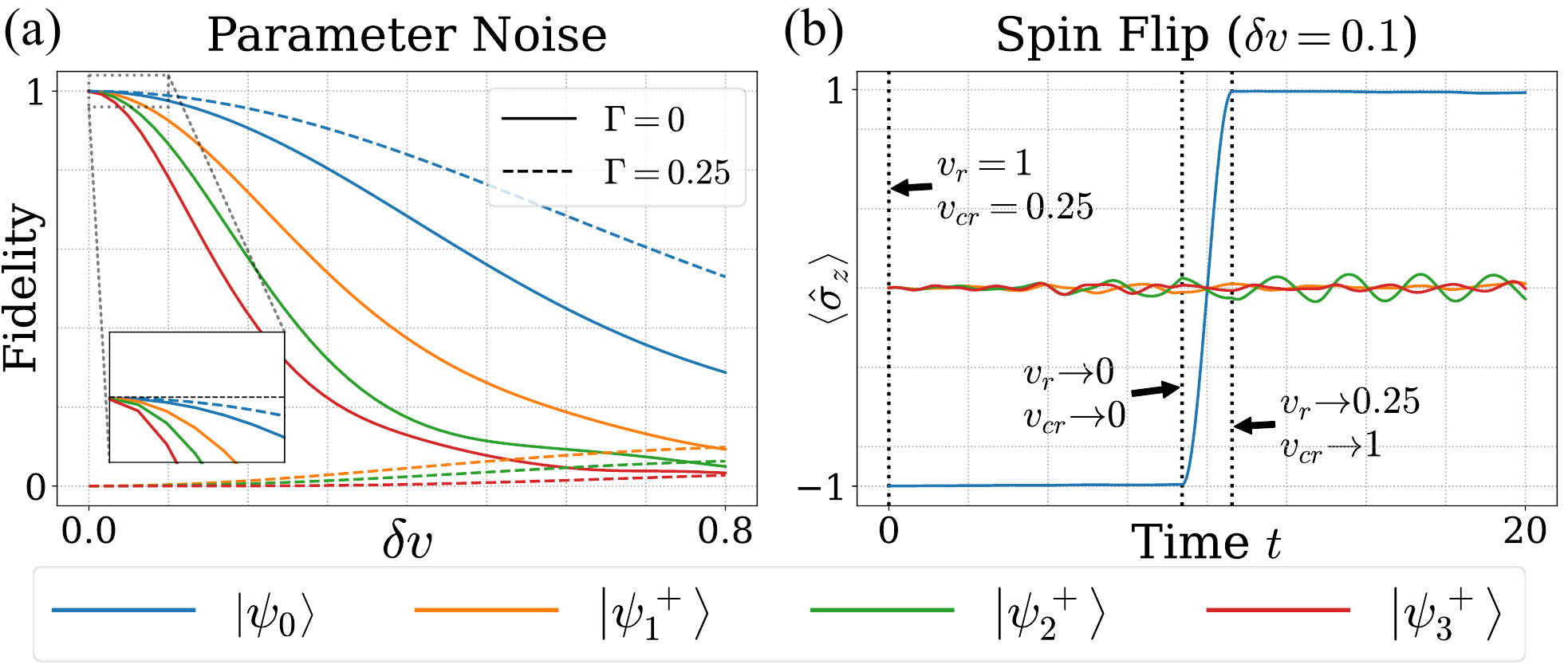}
\caption{
Noise simulation results averaged over 20 Monte Carlo trials at \(w=v_r=1\) and \(v_{cr}=0.25\).
(a) Fidelity between the exact eigenstates and states obtained by adiabatic preparation.
States are adiabatically prepared into \(\ket{\psi_0}\) and \(\left|\psi^+_{1,2,3}\right\rangle\) under noise on \(v_r\) and \(v_{cr}\) with amplitude \(\delta v\). 
Solid and dashed lines show results without (\(\Gamma=0\)) and with (\(\Gamma=0.25\)) dissipation, respectively, for the pumping rate \(\Gamma\).
(b) Spin-flip operation via parameter quenching under fixed noise amplitude \(\delta v=0.1\).
The parameters \(v_r\) and \(v_{cr}\) are set to 0 to induce a spin-flip operation due to \(w\hat{\sigma}_x\). 
After the flip, \(v_r\) and \(v_{cr}\) are restored with their magnitudes also flipped.
}
\label{fig:figure3}
\end{figure}

The well-defined topological quantity in Eq. (6) implies that defect states possess robustness against a specific type of parameter noise---which corresponds to continuous deformation of the system---allowing for practical implications for quantum-state engineering.
For instance, Fig.~\ref{fig:figure3}(a) shows that the zero-energy defect state \(\ket{\psi_0}\) at \(w=v_r=1\) and \(v_{cr}=0.25\) can be generated with high fidelity under small fluctuations in \(v_r\) and \(v_{cr}\).
We initialize the system in the $\alpha=r=0$ state of Eq.~(\ref{eqn:eqn2}) and adiabatically ramp $w$ and $v_{cr}$ in Eq.~(\ref{eqn:eqn3}) from 0 to their target values, with $v_r$ fixed.
During the adiabatic process, uniformly random fluctuations are added to the real and imaginary parts of \(v_r\) and \(v_{cr}\), with amplitudes bounded by \(\delta v\).
In contrast, the adiabatically prepared states \(\ket{\psi_{1,2,3}^+}\) exhibit rapidly decaying fidelities for larger noise amplitudes due to the absence of topological features. 
Interestingly, this robustness is enhanced under the dissipation induced by a conventional Lindblad jump operator \cite{Nielsen_Chuang}, \(\hat{\sigma}_-\), which results in dissipative pumping into the \(\ket{g}\) sublattice~\cite{ion_sdf,reservoir_engineering}, in sharp contrast to fidelity degradation of the other eigenstates (dashed lines for a pumping rate \(\Gamma=0.25\)).
(See Appendix C and the Supplemental Material~\cite{supplement}).

Along with robustness, our winding-number-based analysis, revealing the dependency on the relative magnitude of \(v_r\) and \(v_{cr}\), enables a seemingly contradictory capability: highly selective tunability.
We note that setting \(v_r\) and \(v_{cr}\) to zero and then restoring them with flipped magnitudes induces a spin flip via the \(w\,\hat{\sigma}_x\) interaction.
Figure~\ref{fig:figure3}(b) demonstrates that only the zero-energy defect state \(\ket{\psi_0}\) exhibits a robust, quantized change in \(\langle\hat{\sigma}_z\rangle\), whereas other eigenstates show fluctuations due to parameter noise.
Accordingly, the zero-energy defect state enables a noise-robust, quantized spin control.
Moreover, it allows for direct measurement of the phase-space winding number via oscillator-state tomography; see Supplemental Material~\cite{supplement}.

The physical meaning of the phase-space winding number can be understood by analyzing the geometric phase accumulated along the phase parameter \(\phi\).
As shown in Eqs.~(\ref{eqn:eqn2}) and (\ref{eqn:eqn6}), \(\phi\) characterizes the periodic structure of the eigenstates \( |\psi^{\pm}_n\rangle \) in phase space, and can be interpreted as a quasi-momentum of the SSH-like chain defined on the FSL.  
Accordingly, the accumulated phase difference between the two sublattices over \( \phi \in [0, \pi] \) corresponds to the Zak phase~\cite{Zak_phase} of the SSH-like chain.  
As \(\phi\) varies from 0 to \(\pi\), each Fock state \( |n\rangle \) gains a phase proportional to \(n\) due to the phase-shifting operator \( \hat{R}(\theta) \), forming a helical structure along the \( \phi \)-axis.  
The dressed eigenstates \( |\psi_n^{\pm}\rangle \) in Eq.~(\ref{eqn:eqn2}) thus exhibit a uniform relative phase difference between the \( |A\rangle \) and \( |B\rangle \) sublattices for all \(n\).
As a result, a normalized bulk state \( |\psi\rangle \) acquires a quantized geometric phase in the \( |A\rangle \) sublattice as \( \phi \) varies from 0 to \( \pi \) \cite{supplement}.

As described in Eq. (\ref{eqn:eqn6}), the spin state \( |A\rangle \), which is entangled with the bosonic state \( |n+1\rangle \), is determined by the relative strength between \( |v_r| \) and \( |v_{cr}| \).  
As a result, the sign of the phase difference between the sublattices \( |e\rangle \) and \( |g\rangle \) exhibits a discrete jump at the transition point \( |v_r| = |v_{cr}| \).
To analyze this behavior, we evaluate the accumulated geometric phase difference between the two sublattices for a normalized bulk state \( |\psi\rangle \) as \( \phi \) varies over the interval \([0, \pi]\)~\cite{berry_phase_coherent, squeezed_number_state, supplement}:
\begin{align*}
\Delta \gamma = \frac{i}{\pi\cosh{2r}}\int_{0}^{\pi} d\phi \, \langle \psi | \hat{\sigma}_z | \partial_{\phi} \psi \rangle =
\begin{cases}
1 & |v_r| > |v_{cr}| \\
-1 & |v_r| < |v_{cr}|,
\end{cases}
\end{align*}
which defines the phase-space Zak phase that illustrates a quantum state geometry similar to the winding number in Eq. (\ref{eqn:eqn6}).
The discrete change in \( \Delta \gamma \) at \(|v_r|=|v_{cr}|\) reflects the spin polarization of the SSH-like chain, illustrating the bulk--boundary correspondence in the FSL.  
The winding number, defined along a single loop in \(\phi\)-space, captures the difference in the helical structure between \( |n\rangle \) and \( |n+1\rangle \), with the sign information encoded in the geometric parameters \( R \) and \( C \).

\begin{figure}[t!]
\centering
\includegraphics[width=1.\columnwidth]{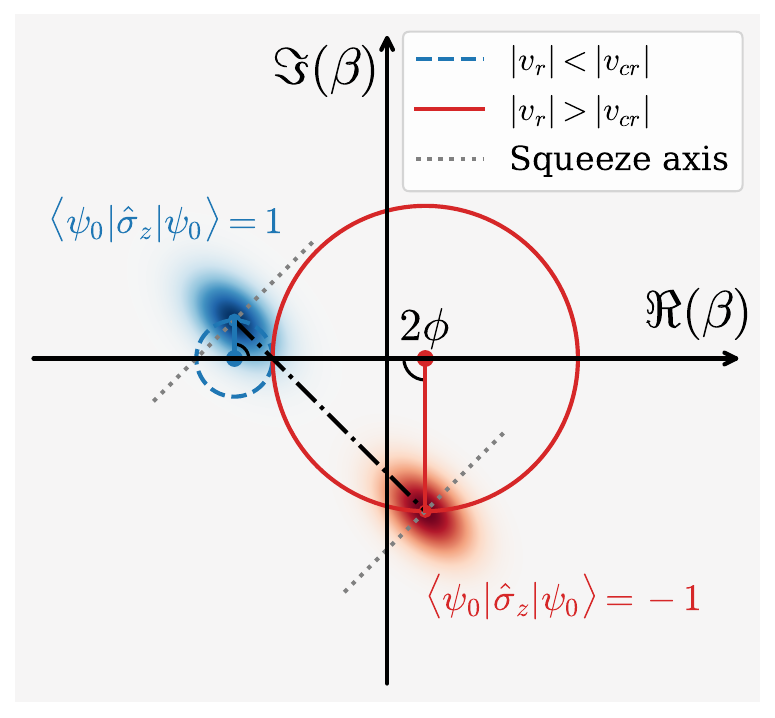}
\caption{
Phase-space trajectories of \(\alpha(\phi)\) at \(w=2.5\) in the frame rotated by \(\phi_{cr}\).
Red solid (blue dashed) trajectory corresponds to \( |v_r| = 1 \), \( |v_{cr}| = 0.25 \) (\( |v_r| = 0.25 \), \( |v_{cr}| = 1 \)). 
The trajectories intersect at \(\phi=0\), and the Wigner functions of the defect states at \(\phi=\pi/4\) are shown for each regime.
Gray dotted lines indicate the squeezing axes, perpendicular to the black dot–dashed lines connecting the phase-space centers \(\alpha(\phi)\) between the two regimes.
}
\label{fig:figure4}
\end{figure}

\textit{Quantum phase-space analysis}---To interpret the observed topological properties in terms of quantum states, we introduce the quantum phase-space illustration of defect states.
The \(\phi\)-dependent energy spectrum, as shown in Eq.~(\ref{eqn:eqn4}) and Appendix A, reflects the critical behavior of defect states emerging in phase space.

Figure~\ref{fig:figure4} illustrates the phase-space trajectories of \(\alpha(\phi)\) for \( |v_r| - |v_{cr}| = \pm 0.75 \), along with the Wigner functions \(\mathcal{W}(\beta) \propto \langle \hat{D}(-\beta) \hat{\Pi} \hat{D}(\beta) \rangle \) of the zero-energy defect states at \( \phi = \pi/4 \), shown in the frame rotated by \(\phi_{cr}\). 
Here, \(\hat{\Pi}\) denotes the parity operator \cite{Wigner_parity}.
The squeezing axes are perpendicular to the black dot-dashed line connecting the centers of the two defect states in different regions. 
As the difference \( |v_{r}| - |v_{cr}| \to 0 \), the defect states become infinitely squeezed along the squeezing axis, and both the center \( C \) and the radius \( R \) of the trajectory diverge. 
At \( \phi = 0 \), the centers of the eigenstates in different phases coincide, as shown in Fig.~\ref{fig:figure4}, and the energy gap in Eq.~(\ref{eqn:eqn4}) closes at the critical point \( \Delta |v| = 0 \). 
In contrast, for \( \phi \ne 0 \), the squeezing axes remain parallel between the two regimes, but a spatial separation between the phase-space centers emerges and reaches a maximum at \( \phi = \pi/2 \). 
This separation diverges as \( \Delta |v| \to 0 \), resulting in a finite energy gap at the critical point. 
This critical behavior underlies the Dirac-cone structure described in Eq.~(\ref{eqn:eqn4}) and Appendix A.

In contrast to classical systems, where topological phenomena emerge from the band structure, the topological features in our system are rooted in its entanglement structure and phase-space characteristics.
The defect state possesses topologically protected spin polarization, and the spin--boson entanglement in Eq.~(\ref{eqn:eqn2}) induces a Zak phase in the single-atom quantum Rabi system.
The phase-space geometry defines the corresponding topological invariant, the phase-space winding number, and the interplay between displacement and squeezing of the eigenstates leads to a conical energy spectrum.
Our results highlight phase-space geometry as a powerful framework for realizing and probing topological phenomena in hybrid quantum spin--oscillator systems~\cite{hybrid_qc_device, hybrid_qc_device2}, where topology arises from intrinsic quantum properties.

\textit{Discussion}---In this work, we have shown that a single-atom quantum Rabi system exhibits topological features characterized by its phase-space geometry.
To address the absence of translational symmetry and the unconventional band structure in FSLs, we introduced a phase-space winding number as a topological invariant that reflects a quantized geometric phase.
This invariant captures the topologically protected spin polarization of defect states in a semi-infinite QRM chain and connects directly to the Zak phase of the FSL.
We further analyzed the \(\phi\)-dependent dynamics in phase space, which led to a conical energy spectrum emerging at the critical point.
Our results show that topological features can arise from the intrinsic quantum nature of a single-atom synthetic lattice, providing a compact and controllable framework for exploring topological phenomena in quantum systems.

In particular, we showed that topological phases, typically characterized by spatial symmetries in real-space lattices, can instead be identified by geometric features in the phase space of a FSL.
This provides a physically motivated, experimentally accessible topological quantity that bridges theoretical studies of semi-infinite FSLs with experimental realizations. 
For example, recent advances in bosonic state tomography \cite{tomography_ion,tomography_circuit,hybrid_ion2} enable direct measurement of the phase-space winding number via tomographic reconstruction~\cite{supplement}. 
We briefly discuss experimental implementations on trapped-ion and superconducting-circuit platforms in Appendix D and in the Supplemental Material~\cite{supplement}.

Another important implication of our results is the separation of spin and bosonic nature in the defect state.
The spin polarization is topologically protected and robust to parameter noise, whereas bosonic properties, such as the occupation \(\langle n\rangle\), designed selective flipping of spin polarization, and the phase-space distribution, remain tunable via system parameters. 
We demonstrated high-fidelity preparation of a nonclassical bosonic state under parameter noise and further enhanced its robustness via engineered dissipation to the \(\ket{g}\) sublattice.
These results open a pathway to the generation of precise nonclassical states even in the presence of noise, with potential applications in quantum metrology~\cite{metrology1,metrology2,metrology3,metrology4,metrology5}.

The Hamiltonian in Eq.~(\ref{eqn:eqn1}) is widely used for gate operations in modern quantum processors~\cite{hybrid_qc_device}, for example to generate spin--motion entanglement in trapped-ion systems~\cite{robust_sdf}.
The same interaction enables multiqubit entanglement via common motional modes~\cite{MS_gate,N_body_gate}, and gates robust to parameter noise have been developed~\cite{robust_gate_ion1,robust_gate_ion2,robust_gate_ion3,robust_gate_ion4,robust_gate_ion5}.
An alternative approach employs bosonic codes with spin-mediated interactions and readout~\cite{hybrid_ion,hybrid_ion2,hybrid_superconducting,hybrid_superconducting2}. 
We investigated distinctive phase-space dynamics and the resulting conical spectrum induced by a coherent spin drive \(w\hat{\sigma}_x\) at the critical point \(|v_r|=|v_{cr}|\). 
These characteristics suggest potential routes to robust gate design and bosonic entangling operations.
We expect our results to extend to topological phenomena arising from interactions between spin-protected defect modes, connecting quantum information science and topological physics through a phase-space perspective.

\textit{Acknowledgments---}Sunkyu Yu was supported by Creative-Pioneering Researchers Program through Seoul National University.
The rest of authors are supported by the Institute for Information \& Communications Technology Planning \& Evaluation (IITP) grant (No.RS-2022-II221040), and the National Research Foundation of Korea (NRF) grant (No. RS-2020-NR049232, No. RS-2024-00442855, No. RS-2024-00413957), all of which are funded by the Korean government (MSIT).

\textit{Data availability---}The simulation codes and data that support the findings of this article are openly available at \cite{git_repo}

\nocite{qutip, circuit_tomography, circuit_tomography2, cavity_tomography, circuit_qed, matt_thesis}
\bibliography{main.bib}

\clearpage
\begin{center}
\textbf{Appendix}
\end{center}

\begin{figure}[h!]
\centering
\includegraphics[width=0.9\columnwidth]{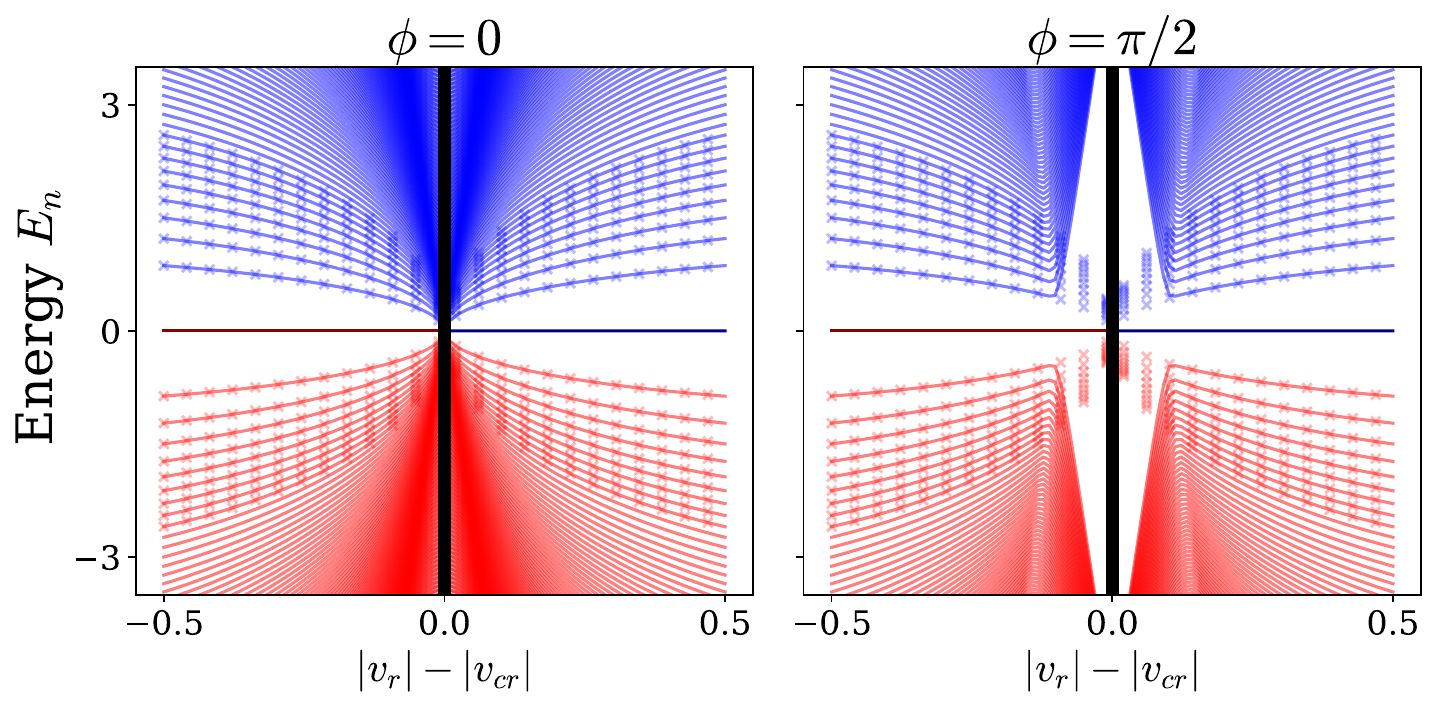}
\caption{
Energy spectrum \( E^{\pm}_n \) of the Hamiltonian in Eq.~(\ref{eqn:eqn1}) as a function of \( |v_r| - |v_{cr}| \), with \( w = 5 \). 
The spectrum is obtained via numerical diagonalization with a Fock-space cutoff \( N_{\max} = 3000 \), for \( \phi = 0 \) and \( \phi = \pi/2 \). 
The singular region \( -0.01 \le |v_r| - |v_{cr}| \le 0.01 \), which is excluded from the calculation, is masked out. 
Each line corresponds to an energy level \( E_n \), and the color indicates its sign. 
Cross markers represent theoretical predictions from Eq.~(\ref{eqn:eqn2}) for \( 1 \le n \le 10 \), showing good agreement with the numerical results, except near the singular region. 
When \(\phi=\pi/2\), the spectrum displays a distinct feature near the point \(|v_r| = |v_{cr}|\), as detailed in the main text.
}
\label{fig:figure5}
\end{figure}

\begin{figure}[b!]
\centering
\includegraphics[width=0.7\columnwidth]{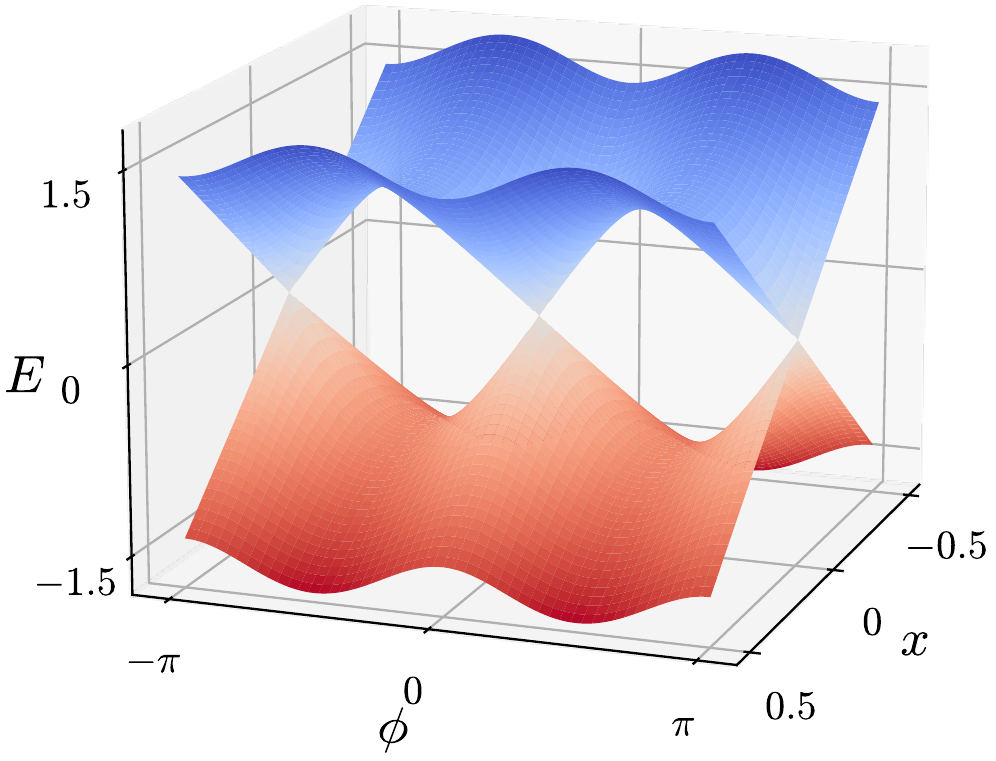}
\caption{
Energy spectrum \(E^{\pm}(x,\phi)\) given in Eq.~(\ref{eqn:eqn4}) for \(w = 1\), \(|v| = 2\). A periodic array of conical Dirac points appears at \((q,\phi) = (0, m\pi)\) for \(m \in \mathbb{Z}\).
}
\label{fig:figure6}
\end{figure}

\textit{Appendix A: Energy Spectrum---}Figure~\ref{fig:figure5} shows the energy spectra obtained via numerical diagonalization of the Hamiltonian in Eq.~(\ref{eqn:eqn1}) for \( \phi = 0 \) and \( \phi = \pi/2 \) under \(|v_{r}|\ne |v_{cr}|\). 
The analytical expression in Eq.~(\ref{eqn:eqn2}), indicated by cross markers, shows excellent agreement with the numerical results. 
Notably, the spectra exhibits qualitatively different behaviors near the critical point \( |v_r| = |v_{cr}| \), depending on the values of \( \phi \), as discussed in the main text.
When \(|v_{r}| = |v_{cr}|\), a conical energy spectrum emerges in a \((1\!+\!1)\)-dimensional space spanned by \(x\) and \(\phi\), as described by Eq.~(\ref{eqn:eqn4}). 
Figure~\ref{fig:figure6} illustrates the resulting periodic Dirac-cone structure in the \((x,\phi)\) space.

\begin{figure}[b!]
\centering
\includegraphics[width=0.8\columnwidth]{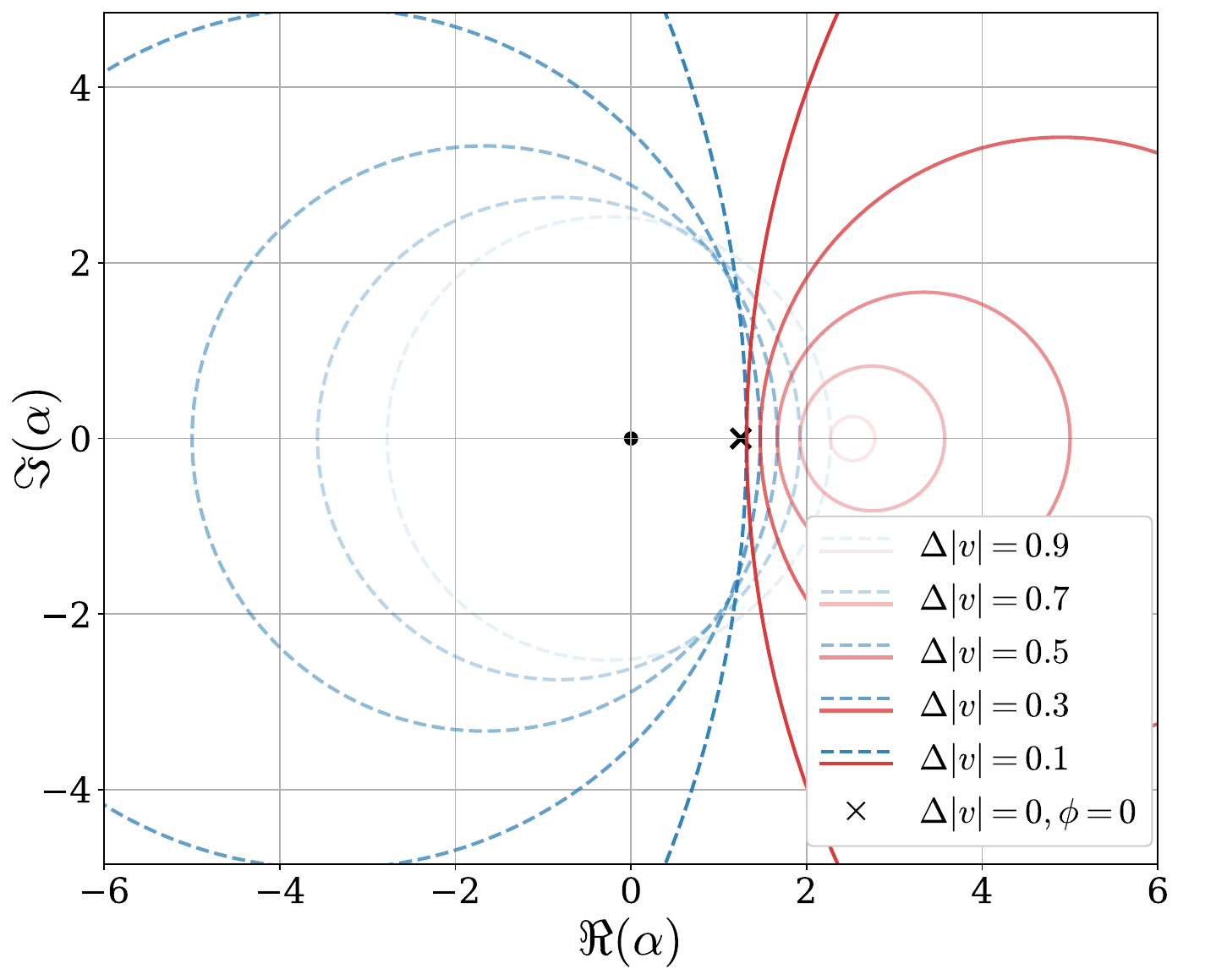}
\caption{
Phase-space trajectories with fixed \(\phi_r\) and \(\phi_{cr}\) swept over \([0, 2\pi]\). Other conditions are the same to Fig.~\ref{fig:figure2}. Red solid (blue dashed) curves correspond to the regime \(|v_r| > |v_{cr}|\) (\(|v_r| < |v_{cr}|\)). 
The geometric properties remain the same, except the condition for whether the trajectory encircles the origin is reversed.
Thus, the two topological phases can still be distinguished by the phase-space winding number.
}
\label{fig:figure7}
\end{figure}

\textit{Appendix B: Phase-Space Trajectory---}All eigenstates in Eq.~(\ref{eqn:eqn2}) are centered at
\begin{equation}
\alpha_r(\phi_r, \phi_{cr}) = \frac{w(v_{cr}^* - v_r)}{|v_r| - |v_{cr}|^2},
\end{equation}
where $\Re\{\alpha_r\} = \sqrt{2} \langle \hat{x} \rangle$ and $\Im\{\alpha_r\} = \sqrt{2} \langle \hat{p} \rangle$. Here, \(\hat{x}\) and \(\hat{p}\) are the position and momentum operators in the ladder basis of \(\hat{a}\) and \(\hat{a}^\dagger\), respectively. The additional degree of freedom \(\phi_{cr}\) can be eliminated by moving to a rotated frame, \(\alpha(\phi) = \alpha_r(\phi) e^{i\phi_{cr}}\). In this frame, the phase-space center becomes
\begin{equation}
\alpha(\phi) = \frac{w(|v_{cr}| - |v_r| e^{i2\phi})}{|v_r|^2 - |v_{cr}|^2}.
\end{equation}
Fixing \(\phi_{cr}\) and varying \(\phi_r\) from \(0\) to \(2\pi\), the trajectories in Fig.~\ref{fig:figure2} correspond to the positions of \(\alpha(\phi)\) over \(\phi \in [0, \pi]\) in the complex \(\alpha\)-plane.  
Since \(\phi_{cr}\) is fixed, we have \(\phi \propto -\theta\) for \(2\theta = \phi_{cr} - \phi_r\), allowing the trajectories to be interpreted as geometric features of the eigenstates as the phase difference between the hopping terms \(v_r\) and \(v_{cr}\) varies from \(0\) to \(2\pi\).  
A single loop of \(\alpha(\phi)\) over \(\phi \in [0, \pi]\) thus captures the helical structure of the eigenstates along the \(\theta\)-axis induced by \(\hat{R}(\theta)\).

Alternatively, one may fix \(\phi_r\) instead of \(\phi_{cr}\). However, this choice does not affect the global property---whether the phase-space trajectories encircle the origin depends only on the relative values of \(|v_r|\) and \(|v_{cr}|\), as shown in Fig.~\ref{fig:figure7}.
That is, the choice of fixed parameter determines only the sign of the relation between \(\phi\) and \(\theta\), while the geometric structure remains unchanged.
Experimentally, the phase-space trajectories can be measured by preparing an eigenstate and tracking its phase-space distribution as a function of \(\phi_r\) or \(\phi_{cr}\), while keeping the other fixed. The choice of which phase parameter to vary depends solely on the experimentalist's preference.
The choice of phase parameter to vary is largely a matter of convenience and has no significant impact on the result.

The phase-space winding number can be calculated as
\begin{align*}
W 
& = \frac{1}{2\pi i} \int_{0}^{\pi} d\phi \, \frac{d}{d\phi} \ln \alpha(\phi) = \frac{1}{2\pi i} \oint_{\mathcal{C}'} \frac{1}{t} \, dt ,
\end{align*}
where $t = |v_{cr}| - |v_r| e^{2i\phi}$. Here, \(\mathcal{C}'\) is the trajectory traced by \(t\) as \(\phi\) varies from \(0\) to \(2\pi\).  
According to the residue theorem, the last integral evaluates to \(2\pi i\) if the contour \(\mathcal{C}'\) encloses the origin, and to zero otherwise. Therefore, the winding number is given by
\begin{align*}
2W = 1 - \mathrm{sgn}(|v_{cr}| - |v_r|).
\end{align*}
If instead we fix \(\phi_r\) and define \(\alpha(\phi) = \alpha_r(\phi) e^{-i\phi_r}\), the contour \(\mathcal{C}'\) becomes the trajectory traced by \(t(\phi) = |v_{cr}| e^{-2i\phi} - |v_r|\), resulting in a winding number:
\begin{align*}
2W = 1 + \mathrm{sgn}(|v_{cr}| - |v_r|).
\end{align*}

\begin{figure}[b!]
\centering
\includegraphics[width=1.\columnwidth]{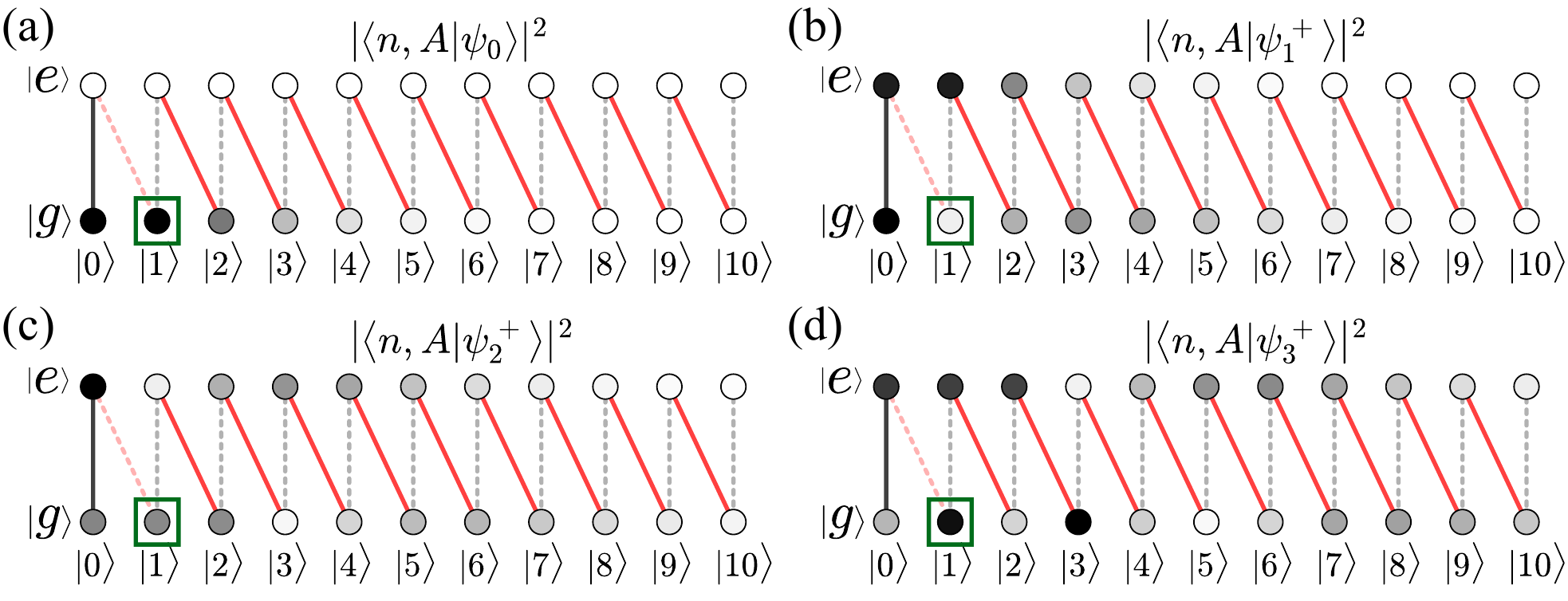}
\caption{
Fock-state populations of eigenstates at \(w=v_r=1\) and \(v_{cr}=0.25\).
For clarity, the weaker hopping \(v_{cr}\) is not shown.
Solid and dashed lines indicate the stronger and weaker of the intra-cell \(w\) and inter-cell \(v_r\) couplings.
Each population \(|\langle n,A|\psi\rangle|^2\) for \(A\in\{e,g\}\) is encoded by the opacity of the circles, with higher opacity indicating larger population.
A domain wall appears at \(n=1\), and the defect \(\ket{1,g}\) is highlighted by a green box.
The zero-energy defect state \(\ket{\psi_0}\) is localized near the defect and resides on the \(\ket{g}\) sublattice.
In contrast, the populations of other eigenstates \(\ket{\psi_k^{+}}\) are more spread over the FSL and occupy both sublattices.
}
\label{fig:figure8}
\end{figure}

\textit{Appendix C: Fock-state Populations---}
This section presents the Fock-state populations of the eigenstates \(\ket{\psi_0}\) and \(\ket{\psi_k^{+}}\) (\(k\in\{1,2,3\}\)) with parameters \(v_r=w=1\) and \(v_{cr}=0.25\) used in the numerical simulations of Fig.~\ref{fig:figure3}.
In this case, a domain wall appears at site \(n=1\), with the defect located at \(\ket{1,g}\).
Figure~\ref{fig:figure8}(a) shows the Fock-state population of the zero-energy defect state \(\ket{\psi_0}\), which is localized near the defect, indicated by a green box. 
The defect state resides on the sublattice \(\ket{g}\), and the topological protection of its spin polarization \(\langle \psi_0 | \hat{\sigma}_z | \psi_0 \rangle\) prevents excitation into the other sublattice \(\ket{e}\) under parameter noise.
In contrast, the other eigenstates \(\ket{\psi_k^{+}}\) exhibit Fock-state populations extended across the FSL, as shown in Fig.~\ref{fig:figure8}(b)–(d). 
These states occupy both sublattices and are therefore vulnerable to parameter noise, since their spin states are not protected.

\textit{Appendix D: Experimental Realization---}The Hamiltonian in Eq.~(\ref{eqn:eqn1}) can be implemented using trapped ions or superconducting circuits, where the coupling strengths \(|v_r|\), \(|v_{cr}|\), and \(w\) correspond to sideband and carrier Rabi frequencies~\cite{dsc_ion, ion_sdf, sideband_superconducting, sideband_superconducting2, sideband_superconducting3, sideband_superconducting4}.
Since the energy gap between \(E_0\) and \(E_{\pm1}\) remains finite unless \(|v_r| = |v_{cr}|\), the zero-energy defect state can be adiabatically prepared by first activating one of the sidebands (\(v_r\) or \(v_{cr}\)) and then gradually increasing both \(w\) and the remaining sideband strength.
Geometric features, such as phase-space trajectories and distributions, can be estimated trough quantum state tomography \cite{hybrid_ion2, tomography_ion, tomography_circuit}.
The main limitation of this adiabatic protocol lies in the trade-off between the adiabatic timescale and the coherence time of the zero-energy state. The adiabatic timescale is determined by the energy gap \(|E_{\pm1} - E_0| = \sqrt{| |v_r| - |v_{cr}| |}\), and thus the coherence time must exceed this timescale to ensure high-fidelity state preparation and measurement.

In trapped-ion systems, typical red- and blue-sideband Rabi frequencies are on the order of tens of kHz, and \(v_r\), \(v_{cr}\), and \(w\) can be independently tuned via radio-frequency (RF) amplitude modulation.  
The motional coherence time near the ground state can approach one second~\cite{motion_coherence_ion}, making trapped-ion platforms promising candidates for probing phase-space characteristics through an adiabatic process.
However, the coherence time is significantly shorter for highly excited Fock states, and the accessible parameter regime may be constrained by the Lamb--Dicke condition~\cite{ion_sdf}, which limits both the maximum achievable sideband coupling strength and the Fock-state population.

In superconducting circuits, sideband Rabi frequencies can reach tens of MHz~\cite{sideband_superconducting4}, allowing for faster adiabatic simulations.  
However, the motional coherence time is relatively shorter than in trapped-ion systems, reaching up to tens of milliseconds~\cite{motion_coherence_circuit, motion_coherence_circuit2}.  
The effective parameter regimes remain similar to trapped-ion systems, as the system's topological features are determined by relative coupling strengths rather than absolute values.

In both platforms, the phase factors \( \phi_r \) and \( \phi_{cr} \) are tunable by controlling the relative phases of RF electronics used for sideband generation, enabling precise control of the desired phase-space geometry.
These systems offer precise control over the drive parameters, making them suitable for the experimental demonstration of single-atom topological features via adiabatic preparation and manipulation of defect states.

\end{document}


\title{Supplemental Material: Phase-Space Topology in a Single-Atom Synthetic Dimension}
\author[1,2,3]{Kyungmin Lee}
\author[4]{Sunkyu Yu}
\author[1,2,3]{Jiyong Kang}
\author[1,2,3]{Seungwoo Yu}
\author[1,2,3]{Wonhyeong Choi}
\author[1,2,3]{Daun Chung}
\author[1,2,3]{Sumin Park}
\author[1,2,3,5]{Taehyun Kim}
\affil[1]{ 
Department of Computer Science and Engineering, Seoul National University, Seoul 08826, Republic of Korea
}
\affil[2]{
Automation and System Research Institute, Seoul National University, Seoul 08826, Republic of Korea}
\affil[3]{
NextQuantum, Seoul National University, Seoul 08826, Republic of Korea}
\affil[4]{
Intelligent Wave Systems Laboratory, Department of Electrical and Computer Engineering, Seoul National University, Seoul 08826, Korea}
\affil[5]{
Institute of Applied Physics, Seoul National University, Seoul 08826, Republic of Korea}

\date{}

\maketitle
\vspace{-1.5cm}
\tableofcontents
\vspace{1cm}

\section{Derivation of Energy Spectrum}
\label{section3}
In this section, we provide the energy spectrum and corresponding eigenstates of the quantum Rabi model Hamiltonian without on-site potentials:
\begin{align}
\hat{H} = w\hat{\sigma}_x + \left( v_r \hat{a}^\dagger \hat{\sigma}_- + v_r^* \hat{a} \hat{\sigma}_+ \right) + \left( v_{cr} \hat{a} \hat{\sigma}_- + v_{cr}^* \hat{a}^\dagger \hat{\sigma}_+ \right),
\label{eqn:eqn1}
\end{align}
where we set \( w \in \mathbb{R} \setminus \{0\} \), \( v_r = |v_r| e^{i\phi_r} \in \mathbb{C} \), and \( v_{cr} = |v_{cr}| e^{i\phi_{cr}} \in \mathbb{C} \). Here, \( v_r^* \) and \( v_{cr}^* \) denote the complex conjugates of \( v_r \) and \( v_{cr} \), respectively. 

\subsection{\texorpdfstring{Asymmetric Case: $|v_r| \ne |v_{cr}|$}{Asymmetric Case}}

Assuming $\left|v_{cr}\right|\neq \left|v_r\right|$, consider the following unitary transformation:
$$
H\to H' = \hat{U}^{\dagger} \hat{H} \hat{U}, \quad \hat{U} = \hat{R}(\theta)\hat{D}(\alpha)\hat{S}(r),
$$ 
$$
\quad \alpha = \frac{w\left( |v_{cr}|e^{-i\phi} - |v_r|e^{i\phi} \right)}{|v_r|^2 - |v_{cr}|^2}, 
\quad {\tanh{2 r}} = \frac{2\left| v_r v_{cr}\right|}{\left|v_r\right|^2 + \left|v_{cr}\right|^2 },
\quad r \in \mathbb{R},
\quad \theta = \frac{\phi_{cr}-\phi_r}{2}, 
\quad \phi = \frac{\phi_{r}+\phi_{cr}}{2},
$$
where \(\hat{D}(\alpha)\), \(\hat{S}\left(r \right)\), and \(\hat{R}(\theta)\) are the displacement, squeeze, and phase-shifting operator defined as \cite{quantum_optics, phase_shift_operator}:
$$
\hat{D}\left(\alpha \right) = \exp{\left( \alpha \hat{a}^{\dagger} - \alpha^* \hat{a} \right)}
\to \hat{D}\left(-\alpha \right)\hat{a}\hat{D}\left(\alpha \right)=\hat{a} + \alpha.
$$ 
$$
\hat{S}\left(r \right)=\exp{\left(\frac{r}{2}\left( \hat{a}^2 - \hat{a}^{\dagger 2} \right)\right)}
\to \hat{S}\left(-r \right)\hat{a}\hat{S}\left( r \right)= \hat{a}\cosh{ r }  - {\hat{a}}^{\dagger}\sinh{r},
$$
$$
\hat{R}\left(\theta \right)={\mathrm{exp} \left(-i\theta {\hat{a}}^{\dagger }\hat{a}\right)\ }
\to \hat{R}\left(-\theta \right)\hat{a}\hat{R}\left(\theta \right)=\hat{a}e^{-i\theta }.
$$ 
Then we obtain:
\begin{align*}
\hat{H}' 
&= \hat{U}^{\dagger} \left( \left( w + v_r \hat{a}^\dagger + v_{cr} \hat{a} \right) \hat{\sigma}_{-} + \left( w + v_r^* \hat{a} + v_{cr}^* \hat{a}^\dagger \right) \hat{\sigma}_{+} \right) \hat{U} \\
&=  e^{i\phi} \left\{ \left( |v_r| \cosh{r} - |v_{cr}| \sinh{r} \right) \hat{a}^\dagger + \left( |v_{cr}| \cosh{r} - |v_r| \sinh{r} \right) \hat{a} \right\} \hat{\sigma}_- + (\text{h.c})
\end{align*}

\noindent If $\left|v_r\right|>\left|v_{cr}\right|$, we have:
$$
\tanh{r} = \left|\frac{v_{cr}}{v_{r}}\right|
\to |v_{cr}|{\cosh{r} } - |v_{r}|{\sinh{r}}=0,
$$
which results in:
\begin{align}
\hat{H}'
=\hat{U}^{\dagger} \hat{H} \hat{U} 
=
\left(|v_{r}|{\cosh{r} } - |v_{cr}| \sinh{r} \right)
\left( {\hat{a}}^{\dagger } 
\hat{\sigma}_{-} + \hat{a} \hat{\sigma}_{+} \right) 
= \sqrt{|v_r|^2 - |v_{cr}|^2} \left( {\hat{a}}^{\dagger } 
\hat{\sigma}_{-} + \hat{a} \hat{\sigma}_{+} \right).
\label{eqn:eqn2}
\end{align}
Now the Hamiltonian \(\hat{H}'\) is in the Jaynes--Cummings (JC) form, and its eigenstates and eigenvalues can be written as:
\begin{align*}
|\psi_0\rangle = \hat{R}(\theta) \hat{D}\left(\alpha \right) \hat{S}(\xi)|0\rangle |g\rangle, \quad 
|\psi_{n+1}\rangle = \hat{R}(\theta) \hat{D}\left(\alpha \right) \hat{S}(\xi) \left(\frac{|n+1\rangle |g\rangle \pm |n\rangle |e\rangle}{\sqrt{2}} \right), 
\end{align*}
\begin{align*}
E_n^{\pm}= \sqrt{\left(|v_r|^2 - |v_{cr}|^2\right)n }, \quad n\in \{0,1,2,\cdots\}.
\end{align*}

\noindent On the other hand, if $\left|v_r\right|<\left|v_{cr}\right|$, we find:
$$
{\tanh{r} }=\left|\frac{v_r}{v_{cr}}\right|
\to |v_r|\cosh{r} - |v_{cr}| \sinh{r} = 0.
$$ 
This yields an anti-JC type Hamiltonian:
\begin{align*}
\hat{H}'
=\hat{U}^{\dagger} \hat{H} \hat{U} 
=
\left(|v_{cr}|{\cosh \xi  }-|v_r|{\sinh \xi }\right)
\left( \hat{a}
\hat{\sigma}_{-} + {\hat{a}}^{\dagger }
\hat{\sigma}_{+} \right)
= 
\sqrt{|v_{cr}|^2 - |v_{r}|^2}  
\left( \hat{a}
\hat{\sigma}_{-} + {\hat{a}}^{\dagger }
\hat{\sigma}_{+} \right).
\end{align*}
In this case, we can find:
\begin{align*}
|\psi_0\rangle = \hat{R}(\theta) \hat{D}\left(\alpha \right) \hat{S}(\xi)|0\rangle |e\rangle, \quad 
|\psi_{n+1}\rangle = \hat{R}(\theta) \hat{D}\left(\alpha \right) \hat{S}(\xi) \left(\frac{|n+1\rangle |e\rangle \pm |n\rangle |g\rangle}{\sqrt{2}} \right), 
\end{align*}
\begin{align*}
E_n^{\pm}= \sqrt{\left(|v_{cr}|^2 - |v_{r}|^2\right)n }, \quad n\in \{0,1,2,\cdots\}.
\end{align*}

\subsection{\texorpdfstring{Symmetric Case: $|v_r| = |v_{cr}|$}{Symmetric Case}}
For the symmetric case $|v_r| = |v_{cr}|$, the Hamiltonian in Eq.~(\ref{eqn:eqn1}) can be written as:
\begin{align*}
\hat{H}
&=\left( w + |v| e^{i\phi_r} \hat{a}^\dagger + |v| e^{i\phi_{cr}} \hat{a} \right) \hat{\sigma}_{-}
+ \left( w + |v| e^{-i\phi_r} \hat{a} + |v| e^{-i\phi_{cr}} \hat{a}^\dagger \right) \hat{\sigma}_{+}
\end{align*}
We consider the following ansatz for the eigenstates:
\[
|\psi\rangle = |f_g\rangle \otimes |g\rangle + |f_e\rangle \otimes |e\rangle, \quad
|f_{g,e}\rangle = \sum_{n=0}^\infty c_n^{(g,e)} |n\rangle.
\]
From the Schrödinger equation, we obtain the following coupled equations:
\begin{align}
\langle g| \hat{H} |\psi\rangle 
&= \left( w + |v|e^{i\phi_r}\hat{a}^\dagger + |v|e^{i\phi_{cr}} \hat{a} \right) |f_e\rangle = E |f_g\rangle, \nonumber \\
\langle e| \hat{H} |\psi\rangle 
&= \left( w + |v|e^{-i\phi_r} \hat{a} + |v|e^{-i\phi_{cr}} \hat{a}^\dagger \right) |f_g\rangle = E |f_e\rangle.
\label{eqn:eqn3}
\end{align}

\subsubsection{Zero-energy mode}
If we set \(E = 0\), the coupled equations in Eq.~(\ref{eqn:eqn3}) decouple as:
\[
\left( w + |v|e^{i\phi_{r}} \hat{a}^\dagger + |v|e^{i\phi_{cr}} \hat{a} \right) |f_e\rangle = 0, \quad
\left( w +|v|e^{-i\phi_{r}}  \hat{a} + |v|e^{-i\phi_{cr}} \hat{a}^\dagger \right) |f_g\rangle = 0.
\]
We define the operator
\[
\hat{K} = w + |v| e^{i\phi_r} \hat{a}^\dagger + |v| e^{i\phi_{cr}} \hat{a}
       = w + |v| e^{i\phi} \left( e^{-i \theta} \hat{a}^\dagger + e^{i\theta} \hat{a} \right),
\]
where \(\phi = \frac{\phi_r + \phi_{cr}}{2}\) and \(\theta = \frac{\phi_{cr} - \phi_{r}}{2}\). The two equations can then be written as
\[
\hat{K} |f_e\rangle = 0, \quad \hat{K}^\dagger |f_g\rangle = 0.
\]
The operator \(\hat{K}\) can be transformed as:
\[
\hat{K} \to \hat{R}(-\theta)\hat{K}\hat{R}(\theta) = w + |v| e^{i\phi} (\hat{a}^\dagger + \hat{a}) = w + \sqrt{2} |v| e^{i\phi} \hat{x},
\]
where \(\hat{x} = (\hat{a} + \hat{a}^\dagger)/\sqrt{2}\) is the position operator. Therefore, satisfying \(\hat{K}|f_e\rangle = 0\) and \(\hat{K}^\dagger|f_g\rangle = 0\) requires
\[
|f_g\rangle= |f_e\rangle = \hat{R}(\theta) |q\rangle,
\]
\[
\hat{K}|f_e\rangle
=\hat{R}(\theta)\left(w + \sqrt{2} |v| e^{i\phi} \hat{x}\right)\hat{R}(-\theta) |f_e\rangle
=\left(w + \sqrt{2} |v| e^{i\phi} x\right) |f_e\rangle
=0 \to x=-\frac{w}{\sqrt{2}|v|} e^{-i\phi},
\]
\[
\hat{K}^\dagger|f_g\rangle
=\hat{R}(\theta)\left(w + \sqrt{2} |v| e^{i\phi} \hat{x}\right)\hat{R}(-\theta) |f_g\rangle
=\left(w + \sqrt{2} |v| e^{-i\phi} x\right) |f_g\rangle
=0 
\to x=-\frac{w}{\sqrt{2}|v|} e^{i\phi},
\]
where \(|q\rangle\) is an eigenstate of the position operator \(\hat{q}|q\rangle=q|q\rangle\). This condition can only be satisfied when \(\phi = m\pi\) for \(m \in \mathbb{Z}\). Physically, this corresponds to an infinitely squeezed state with vanishing position uncertainty and divergent momentum uncertainty, which cannot be realized in any physical system. Mathematically, \(|q\rangle\) is not normalizable and lies outside the Hilbert space, satisfying \(\langle q | q \rangle = \delta(0)\). Therefore, no normalizable zero-energy state exists when \(|v_r| = |v_{cr}|\).

\subsubsection{Nonzero energy modes}

We start from the coupled equation in Eq.~(\ref{eqn:eqn3}):
$$\left(w+\left|v\right|e^{i{\phi }_r}{\hat{a}}^{\dagger }+\left|v\right|e^{i{\phi }_{cr}}\hat{a}\right)\left|f_e\right\rangle =E\left|f_g\right\rangle, $$ 
$$\left(w+\left|v\right|e^{-i{\phi }_r}\hat{a}+\left|v\right|e^{-i{\phi }_{cr}}{\hat{a}}^{\dagger }\right)\left|f_g\right\rangle =E\left|f_e\right\rangle, $$ 
For \(E\ne0\), we have:
\begin{align*}
E^2\left|f_g\right\rangle &=\left(w+\left|v\right|e^{i{\phi }_r}{\hat{a}}^{\dagger }+\left|v\right|e^{i{\phi }_{cr}}\hat{a}\right)\left(w+\left|v\right|e^{-i{\phi }_r}\hat{a}+\left|v\right|e^{-i{\phi }_{cr}}{\hat{a}}^{\dagger }\right)\left|f_g\right\rangle \\
&=\left(w^2+{\left|v\right|}^2+w\left|v\right|A\hat{a}+w\left|v\right|A^*{\hat{a}}^{\dagger }+{\left|v\right|}^2B\hat{a}^{\dagger 2}+{\left|v\right|}^2B^*{\hat{a}}^2+2{\left|v\right|}^2{\hat{a}}^{\dagger }\hat{a}\right)\left|f_g\right\rangle ,
\end{align*}
$$A=e^{-i{\phi }_r}+e^{i{\phi }_{cr}},\ \ \ B=e^{i{\phi }_r-i{\phi }_{cr}}.$$ 
By applying the transformation:
$$
\hat{a}\to\hat{D}\left(-\alpha \right)\hat{a}\hat{D}\left( \alpha \right)=\hat{a}+\alpha, \ \ \ \ {\hat{a}}^{\dagger }=\hat{D}\left(-\alpha \right){\hat{a}}^{\dagger }\hat{D}\left(\alpha \right)={\hat{a}}^{\dagger } + {\alpha }^*, \quad
\alpha =-\frac{w}{2\left|v\right|}e^{i{\phi }_r},
$$ 
we obtain:
\begin{align*}
&E^2\left|f_g\right\rangle=\hat{U}_\alpha \left({\left|v\right|}^2B\hat{a}^{\dagger 2}+{\left|v\right|}^2B^*{\hat{a}}^2+{\left|v\right|}^2\left({\hat{a}}^{\dagger }\hat{a}+\hat{a}{\hat{a}}^{\dagger }\right)\right)\hat{U}_\alpha^\dagger\left|f_g\right\rangle + C, \\
& C = w^2+2{\left|v\right|}^2{\left|\alpha \right|}^2+w\left|v\right|\left(A\alpha +A^*{\alpha }^*\right)+{\left|v\right|}^2\left(B{\alpha}^{* 2}+B^*{\alpha }^2\right).
\end{align*}
where \(\hat{U}_\alpha=\hat{D}\left(\alpha \right)\). Excepting the constant term $C$, we have:
$${\left|v\right|}^2 \hat{U}_\alpha \left(B\hat{a}^{\dagger 2}+B^*{\hat{a}}^2+\left({\hat{a}}^{\dagger }\hat{a}+\hat{a}{\hat{a}}^{\dagger }\right)\right)\hat{U}_\alpha^\dagger.$$ 
Again, we apply a transformation:
$$
{\hat{a}}^{\dagger } \to \hat{R}\left(-\theta\right) \hat{a}^{\dagger} \hat{R}\left(\theta\right) = {\hat{a}}^{\dagger }e^{i{\theta }},\quad
\hat{a}= \hat{R}\left(-\theta\right) \hat{a} \hat{R}\left(\theta\right)= \hat{a}e^{-i{\theta }}, \quad{\theta }=\frac{{\phi }_{cr}-{\phi }_{r}}{2}.
$$ 
Then we have:
$${\left|v\right|}^2 \hat{U} \left(\hat{a}^{\dagger 2}+{\hat{a}}^2+{\hat{a}}^{\dagger }\hat{a}+\hat{a}{\hat{a}}^{\dagger }\right) \hat{U}^\dagger
={\left|v\right|}^2 \hat{U}{\left({\hat{a}}^{\dagger }+\hat{a}\right)}^2 \hat{U}^{\dagger}, \quad
\hat{U}=\hat{R}(\theta)\hat{D}(\alpha)$$ 
\noindent Note that the constant term can be recast in the following form:
\begin{align*}
&w^2+2{\left|v\right|}^2{\left|\alpha \right|}^2+w\left|v\right|\left(A\alpha +A^*{\alpha }^*\right)+{\left|v\right|}^2\left(B{\alpha}^{* 2}+B^*{\alpha }^2\right) =\frac{w^2}{2} \left(1 - \frac{1}{2}\left(e^{-\left(i{\phi }_r+i{\phi }_{cr}\right)} + e^{i{\phi }_r+i{\phi }_{cr}}\right) \right).
\end{align*} 
Hence, the original equation can be written as:
$$E^2\left|f_g\right\rangle =\hat{U}\left( \frac{w^2}{2} - \frac{w^2}{4}\left(e^{-\left(i{\phi }_r+i{\phi }_{cr}\right)}+e^{i{\phi }_r+i{\phi }_{cr}} \right) + {\left|v\right|}^2{\left({\hat{a}}^{\dagger }+\hat{a}\right)}^2\right)\hat{U}^\dagger\left|f_g\right\rangle.$$ 
Using the relation ${\hat{a}}^{\dagger }+\hat{a}=\sqrt{2}{\hat{x}}$, with $\hat{x}$ representing the position operator in the ladder basis \(\hat{a}\) and \(\hat{a}^\dagger\), the above equation can be simplified to a form expressed solely in terms of $\hat{x}$:
\begin{align*}
E^2{\left|f_g\right\rangle }
&=\hat{U}\left(\frac{w^2}{2} - \frac{w^2}{4}\left(e^{-\left(i{\phi }_r+i{\phi }_{cr}\right)}+e^{i{\phi }_r+i{\phi }_{cr}} \right) + 2{\left|v\right|}^2{\hat{x}}^2\right)\hat{U}^\dagger{\left|f_g\right\rangle }.
\end{align*}
Therefore, by setting \(|f_g\rangle = \hat{U}|x\rangle\), we have:
\begin{align*}
E^2{\left|x\right\rangle }
&=\left(\frac{w^2}{2}\left(1-{\mathrm{cos} \left({\phi }_r+{\phi }_{cr}\right)\ }\right)+2{\left|v\right|}^2 x^2\right){\left|f_g\right\rangle },
\end{align*}
where $\left|x\right\rangle$ denotes the eigenstate of the position operator $\hat{x}$, $\hat{x}|x\rangle = x|x\rangle$.
Therefore, the corresponding energy spectrum in the continuous position basis associated with $x$ is given by:
\begin{align*}
E\left(x, \phi\right)=\pm \sqrt{\left(\frac{w^2}{2}\left(1-{\cos 2\phi\ }\right)+2{\left|v\right|}^2 x^2\right)},
\end{align*}
where $\phi=({\phi }_r+{\phi }_{cr})/2$. The corresponding eigenstates have the form:
$$\left|\psi_x^{\pm} \right\rangle \propto \hat{U}\left| x \right\rangle \left(\left| g \right\rangle \pm \left| e \right\rangle \right)$$
This implies that as we approach the critical condition $|v_r| = |v_{cr}|$, the energy gap decreases, and the spectrum eventually becomes continuous.

\noindent The bosonic state \(|q\rangle =\hat{U}\left| x \right\rangle\) can be described in the \((\hat{x}, \hat{p})\) basis, where \(\hat{x} = \frac{1}{\sqrt{2}}({\hat{a}}^{\dagger} + \hat{a})\) and \(\hat{p} = \frac{i}{\sqrt{2}}({\hat{a}}^{\dagger} - \hat{a})\) are the position and momentum operators of the bosonic mode \(\hat{a}\), respectively.
\begin{align*}
\hat{x} =\frac{1}{\sqrt{2}}\left({\hat{a}}^{\dagger }+\hat{a}\right) \to \hat{q}
&= \hat{R}(-\theta)\hat{D}(-\alpha)\hat{x}\hat{D}(\alpha)\hat{R}(\theta)  
=\frac{1}{\sqrt{2}}\left(e^{i\theta}{\hat{a}}^{\dagger }+e^{-i\theta}\hat{a}+2Re\left\{ \alpha \right\}\right) \\
&=\frac{1}{\sqrt{2}}\left(e^{i\frac{\left({\phi }_{cr}-{\phi }_{r}\right)}{2}}{\hat{a}}^{\dagger }+e^{-i\frac{\left({\phi }_{cr}-{\phi }_{r}\right)}{2}}\hat{a} - \frac{w}{\left|v\right|} \cos{\phi_r} \right).
\end{align*}
From the relation:
$$\hat{a}=\frac{1}{\sqrt{2}}\left({\hat{x}}+i{\hat{p}}\right),\ \ \ {\hat{a}}^{\dagger }=\frac{1}{\sqrt{2}}\left({\hat{x}}-i{\hat{p}}\right),$$ 
we get the expression of $\hat{q}$ in terms of $\hat{x}$ and $\hat{p}$:
$${\hat{q}}
=
{\mathrm{cos} \left(\frac{{\phi }_{cr}-{\phi }_r}{2}\right)\ }{\hat{x}}
+{\mathrm{sin} \left(\frac{{\phi }_{cr}-{\phi }_r}{2}\right)\ }{\hat{p}}
+\frac{w}{\sqrt{2}\left|v\right|}\cos{\phi_r}
.$$ 
One can also find the expression for the conjugate operator $\hat{q}_p$ such that $[\hat{q},\hat{q}_p]=[\hat{x},\hat{p}]$:
$${\hat{q}_p}
=
{\mathrm{cos} \left(\frac{{\phi }_{cr}-{\phi }_r}{2}\right)\ }{\hat{p}}
-{\mathrm{sin} \left(\frac{{\phi }_{cr}-{\phi }_r}{2}\right)\ }{\hat{x}}
-\frac{w}{\sqrt{2}\left|v\right|}\sin{\phi_r}
.$$ 
The above relation can be expressed using the displacement and phase-shifting operators:
$$\hat{q} = \hat{R}(-\theta) \hat{D}(-\alpha) \hat{x} \hat{D}(\alpha) \hat{R}(\theta), \quad 
\hat{q}_p = \hat{R}(-\theta) \hat{D}(-\alpha) \hat{p} \hat{D}(\alpha) \hat{R}(\theta)$$
$$\alpha = -\frac{w}{2|v|}e^{i\phi_r}, \quad \theta=\frac{\phi_{cr}-\phi_r}{2}.$$

\section{Phase-Space Zak Phase}
Consider the relations:
\begin{align*}
&|\psi_{n+1}^{\pm}\rangle = \hat{R}(\theta) \hat{D}(\alpha) \hat{S}(r) \left( \frac{|n{+}1\rangle |A\rangle \pm e^{i\phi_n} |n\rangle |B\rangle}{\sqrt{2}} \right), \\
&|\psi_0\rangle = \hat{R}(\theta) \hat{D}(\alpha) \hat{S}(r) |0\rangle |A\rangle, \quad E_n^{\pm} = \pm \sqrt{ \left| \left( \left| v_r \right| - \left|v_{cr} \right| \right) n \right| },
\end{align*}
with
\begin{align*}
\alpha &= \frac{w (|v_{cr}|e^{-i\phi} - |v_r|e^{i\phi})}{ |v_r|^2 - |v_{cr}|^2 }, \quad 
\tanh(2r) = \frac{2 |v_r v_{cr}|}{ |v_r|^2 + |v_{cr}|^2 }, \\
&\theta =\frac{\phi_{cr}-\phi_r}{2}, \quad 
\phi=\frac{\phi_r + \phi_{cr}}{2}.
\end{align*}

\noindent Following the condition used in the main text, we fix $\phi_{cr}$ and compute the Berry phase of the state $|\phi_n\rangle=\hat{R}(\theta) \hat{D}(\alpha) \hat{S}(r)|n\rangle$ along the path of $\phi$. Using the relation $\theta = -\phi + \phi_{cr}$, we obtain:
\begin{align*}
&\int_{0}^{\pi} d\phi  \langle \phi_n|\partial_\phi \phi_n  \rangle \\
&=\int_{0}^{\pi} d\phi \, \langle n | \left(\hat{S}(-r) \hat{D}(-\alpha)\hat{R}(\phi-\phi_{cr})\right) \partial_{\phi} \left(\hat{R}(-\phi+\phi_{cr}) \hat{D}(\alpha) \hat{S}(r)\right)|n \rangle
\end{align*}

\noindent Note that the constant phase shift $\phi_{cr}$ can be removed without loss of generality using the identity $\hat{I} = \hat{R}(\phi_{cr}) \hat{R}(-\phi_{cr})$, where $\hat{I}$ denotes the identity operator. Then we have:
\begin{align*}
&\int_{0}^{\pi} d\phi \, \langle n | \hat{S}(-r) \hat{D}(-\alpha)\hat{R}(\phi) \left(\partial_{\phi} \hat{R}(-\phi)\right) \hat{D}(\alpha) \hat{S}(r)|n \rangle
+ \int_{0}^{\pi} d\phi \, \langle n | \hat{S}(-r) \hat{D}(-\alpha) \left(\partial_{\phi} \hat{D}(\alpha) \right) \hat{S}(r)|n \rangle.
\end{align*}

\noindent The second term yields a constant phase \(\Theta(\alpha,r)\), independent of \(n\)~\cite{berry_phase_coherent}. The first term gives:
\begin{align*}
&\int_{0}^{\pi} d\phi \, \langle n | \hat{S}(-r) \hat{D}(-\alpha)\hat{R}(\phi) \left(\partial_{\phi} \hat{R}(-\phi)\right) \hat{D}(\alpha) \hat{S}(r)|n \rangle \\
&=i\int_{0}^{\pi} d\phi \, \langle n | \hat{S}(-r) \hat{D}(-\alpha) \left( \hat{a}^\dagger \hat{a} \right) \hat{D}(\alpha) \hat{S}(r)|n \rangle \\
&=i\int_{0}^{\pi} d\phi \, \langle n | \hat{S}(-r) (\hat{a}^\dagger + \alpha^*)(\hat{a} + \alpha) \hat{S}(r)|n \rangle \\
&=i\int_{0}^{\pi} d\phi\, \langle n | \hat{S}(-r) (\hat{a}^\dagger \hat{a} + \alpha^* \hat{a} + \alpha \hat{a}^\dagger + |\alpha|^2) \hat{S}(r)|n \rangle \\
&=i\int_{0}^{\pi} d\phi \, \langle n | (\hat{n}\cosh^2 r + (\hat{n}+1) \sinh^2 r  + |\alpha|^2) |n \rangle \\
&=i\left( n \pi \cosh (2r) + \pi \sinh^2 \alpha +  + \pi |\alpha|^2\right) 
\end{align*}

\noindent Therefore, the spin-resolved Zak phase for the eigenstate \(|\psi_n^{\pm}\rangle\) for \(|v_r| > |v_{cr}|\) is
\begin{align*}
\Delta \gamma_n 
&= \frac{i}{\pi\cosh{2r}}\int_{0}^{\pi} d\phi \, \langle \psi_n^{\pm} | \hat{\sigma}_z | \partial_{\theta} \psi_n^{\pm} \rangle \\
&=\frac{i}{\pi\cosh{2r}}\int_{0}^{\pi} d\phi \  \left\{ \langle n | \hat{U}^{\dagger} \partial_{\phi} \hat{U} |n\rangle - \langle n+1 | \hat{U}^{\dagger} \partial_{\phi} \hat{U} |n+1\rangle \right\} \\
&=\frac{i}{\pi\cosh{2r}}\times \left(-i \pi \cosh(2r)\right) =1
\end{align*}
where \(\hat{U}=\hat{R}(\theta) \hat{D}(\alpha) \hat{S}(r)\). For \(|v_r| < |v_{cr}|\), we have \(\Delta \gamma_n = 1\) since the spin states coupled to \(|n\rangle\) and \(|n+1\rangle\) are reversed. 
Therefore, the Zak phase \(\Delta \gamma_n\) of a normalized state in the bulk regime, \(|\psi\rangle = \sum_{n} \left( c_n^{+}|\psi_n^{+} + c_n^{-}|\psi_n^{-}\rangle\right)\) with \(\sum_n (|c_n^+|^2+|c_n^-|^2)=1\), is quantized as
\begin{align*}
\Delta \gamma =
\begin{cases}
1 & \text{if } |v_r| > |v_{cr}|, \\
-1 & \text{if } |v_r| < |v_{cr}|.
\end{cases}
\end{align*}

\noindent If we instead fix \(\phi_r\), the sign of \(\Delta \gamma\) is reversed since \(\theta \propto \phi\).

\clearpage

\section{Real-Space Topological Marker for Fock-State Lattices}

Here, we demonstrate an example case where a real-space topooplogical marker is applied to estimate the topological property of the SSH-like chain in a semi-infinite Fock-state lattice. 
We consider a topological marker for the Fock-state lattice, based on the real-space marker introduced for the 1D SSH model in Ref.~\cite{topological_marker}. 
\begin{align}
\hat{C}(n) = \hat{\sigma}_z \left[ \hat{Q} \hat{n} \hat{P} + \hat{P} \hat{n} \hat{Q} \right], \nonumber
\end{align}
where \( \hat{P} \) and \( \hat{Q} \) are projection operators onto the positive- and negative-energy eigenstates, respectively. 

\begin{figure}[b!]
\centering
\begin{overpic}[width=.825\linewidth]{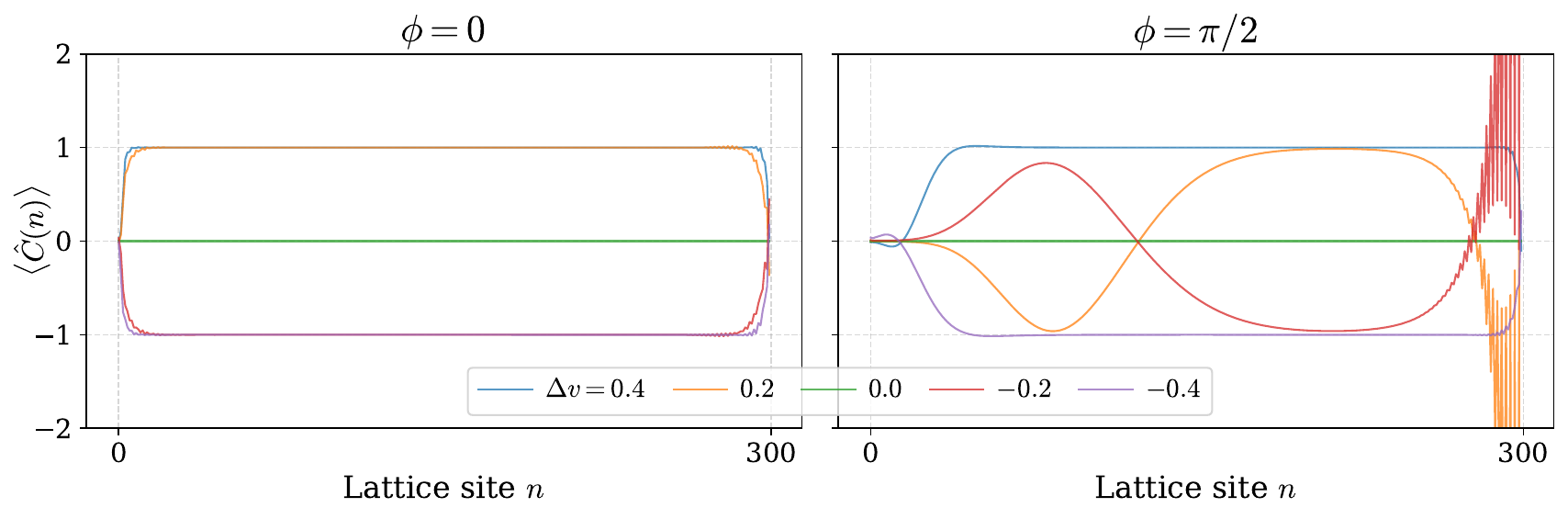}
\put(0,31){(a)}
\end{overpic}
\hspace{0pt}
\begin{overpic}[width=.825\linewidth]{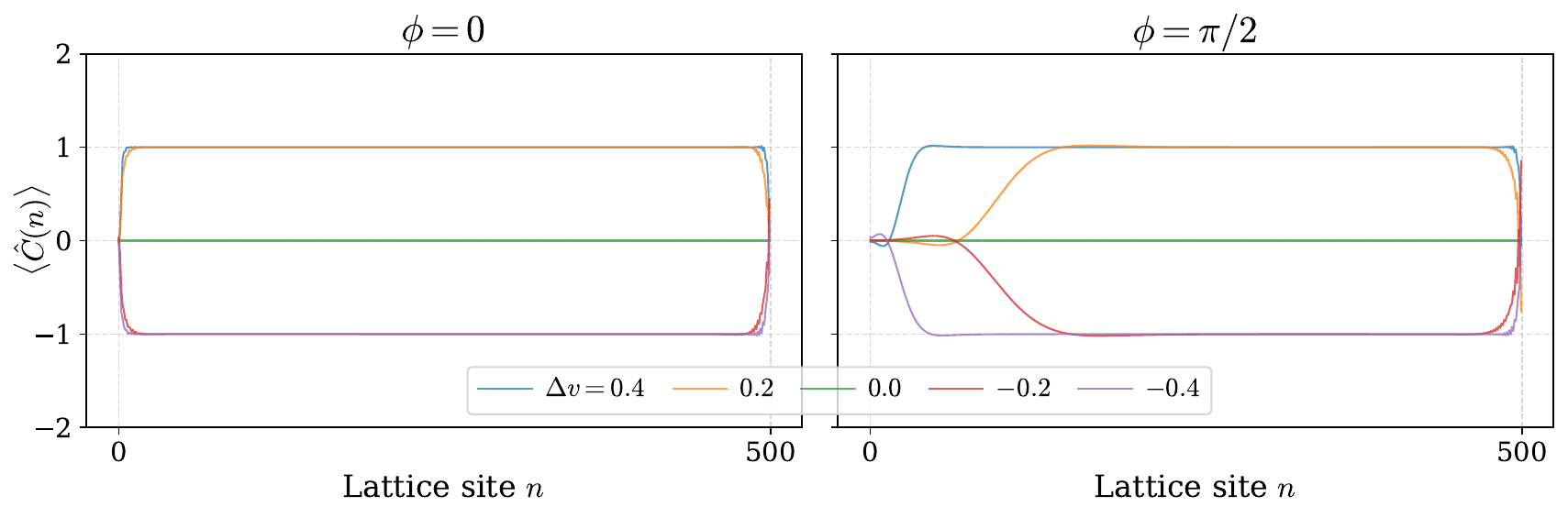}
\put(0,31){(b)}
\end{overpic}
\caption{
Local marker values at each lattice site \(n\) for various \(\Delta v \equiv |v_r| - |v_{cr}|\). 
We use \(w=2\) with (a) Fock-space dimension \(N=300\) and (b) \(N=500\). 
For \(\phi=0\), the marker is flat across the bulk region. 
For \(\phi=\pi/2\), fluctuations appear near the edges for \(N=300\) due to finite-size effects, but are further suppressed for \(N=500\).
}
\label{fig:figures1}
\end{figure}

Figure~\ref{fig:figures1} shows local marker values on lattices truncated at (a) \(N=300\) and (b) \(N=500\).
For \(\phi \equiv (\phi_r+\phi_{cr})/2 = 0\), the local markers in the bulk exhibit a discrete jump at the critical point \(\Delta v \equiv |v_r|-|v_{cr}| = 0\) and are otherwise stable.
By contrast, for \(\phi=\pi/2\) they display fluctuations around the edges due to finite-size effects. 
As the size of the Fock-space $N$ increases, these fluctuations are suppressed, as shown in Fig.~\ref{fig:figures1}(b). 
However, since a truly infinite-dimensional system is hardly realizable, residual fluctuations may persist.
Consequently, fluctuations near the \(n=0\) edge require evaluating the marker over sites \(n > N_{\mathrm{cut}}\) to obtain a reliable estimate, where the cutoff \(N_{\mathrm{cut}}\) is chosen beyond the fluctuation-affected region and depends on the parameters.
In practice, this requirement can be demanding; if \(N_{\mathrm{cut}}\) is chosen too small, the marker becomes sensitive to parameter noise, compromising its reliability.

\begin{figure}[t!]
\centering
\begin{overpic}[width=.825\linewidth]{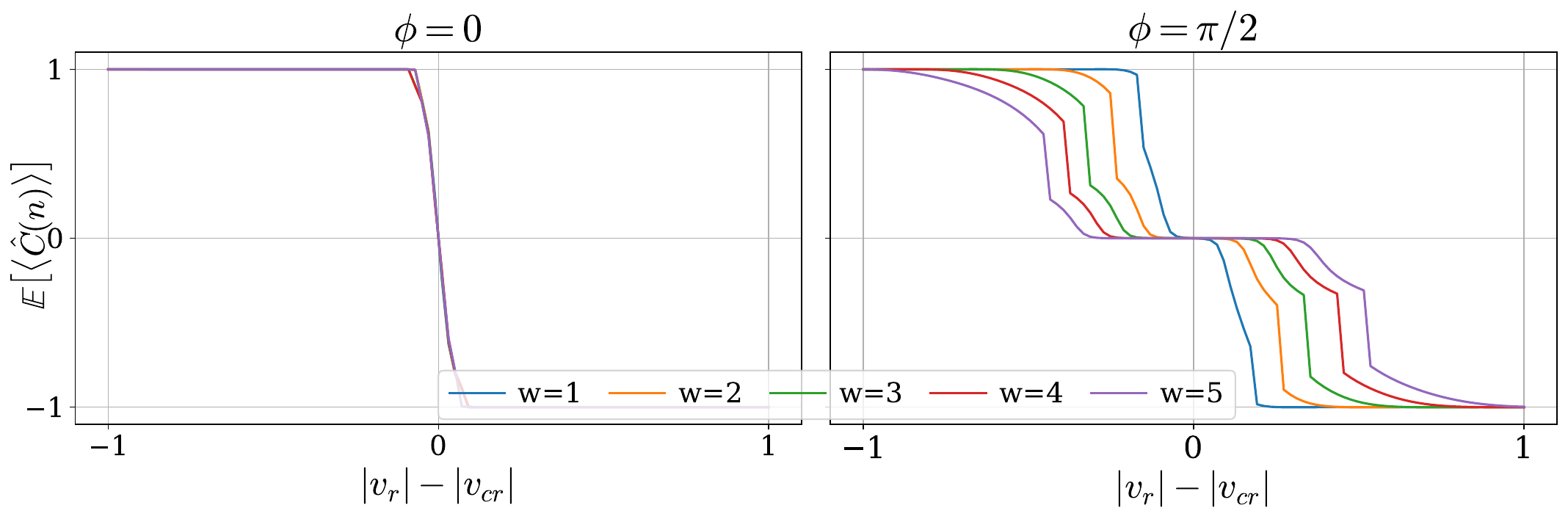}
\put(0,31){(a)}
\end{overpic}
\hspace{0pt}
\begin{overpic}[width=.825\linewidth]{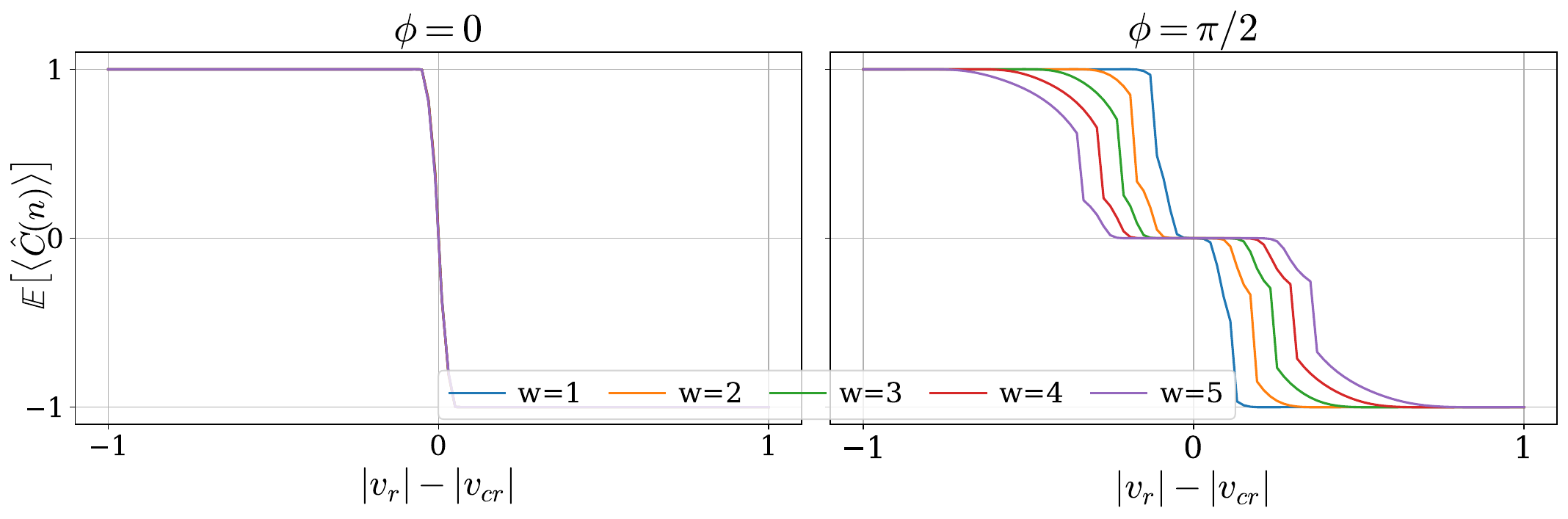}
\put(0,31){(b)}
\end{overpic}
\caption{
Global marker values \( \mathbb{E}[\hat{C}(n)] \) as a function of \( \Delta v \) for \( \phi = 0 \) and \( \phi = \pi/2 \), shown for various values of \( w \).  
The left panel demonstrates that the global marker remains invariant with respect to \( w \) for \( \phi = 0 \),  
while the right panel shows that, for \( \phi = \pi/2 \), the marker exhibits \( w \)-dependence due to the \(\phi\)-dependent distribution of eigenstates in the Fock-state basis.
}
\label{fig:figures2}
\end{figure}

\begin{figure}[b!]
\centering
\begin{tikzpicture}
\node at (0,0) {\includegraphics[width=.3\columnwidth]{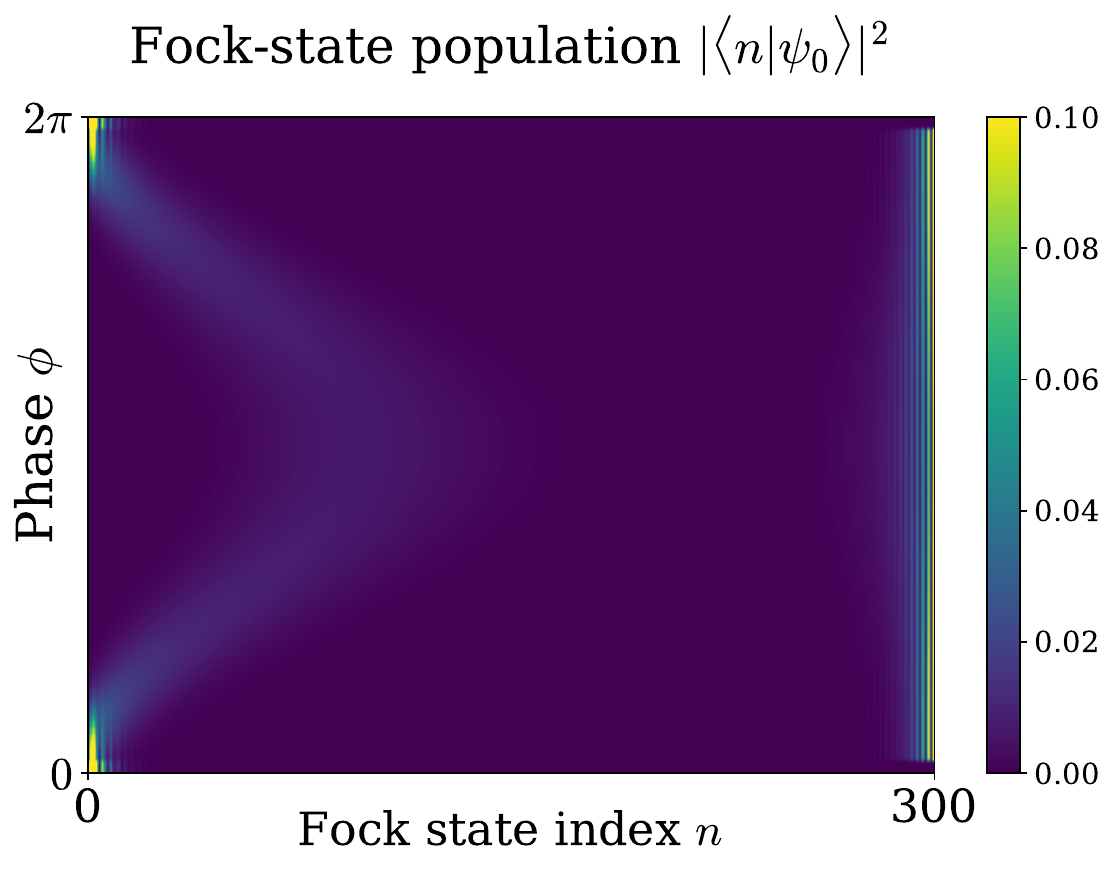}};
\node at (-2.2, 1.7) {\small (a)};
\end{tikzpicture}
\hfill
\begin{tikzpicture}
\node at (0,0) {\includegraphics[width=.3\columnwidth]{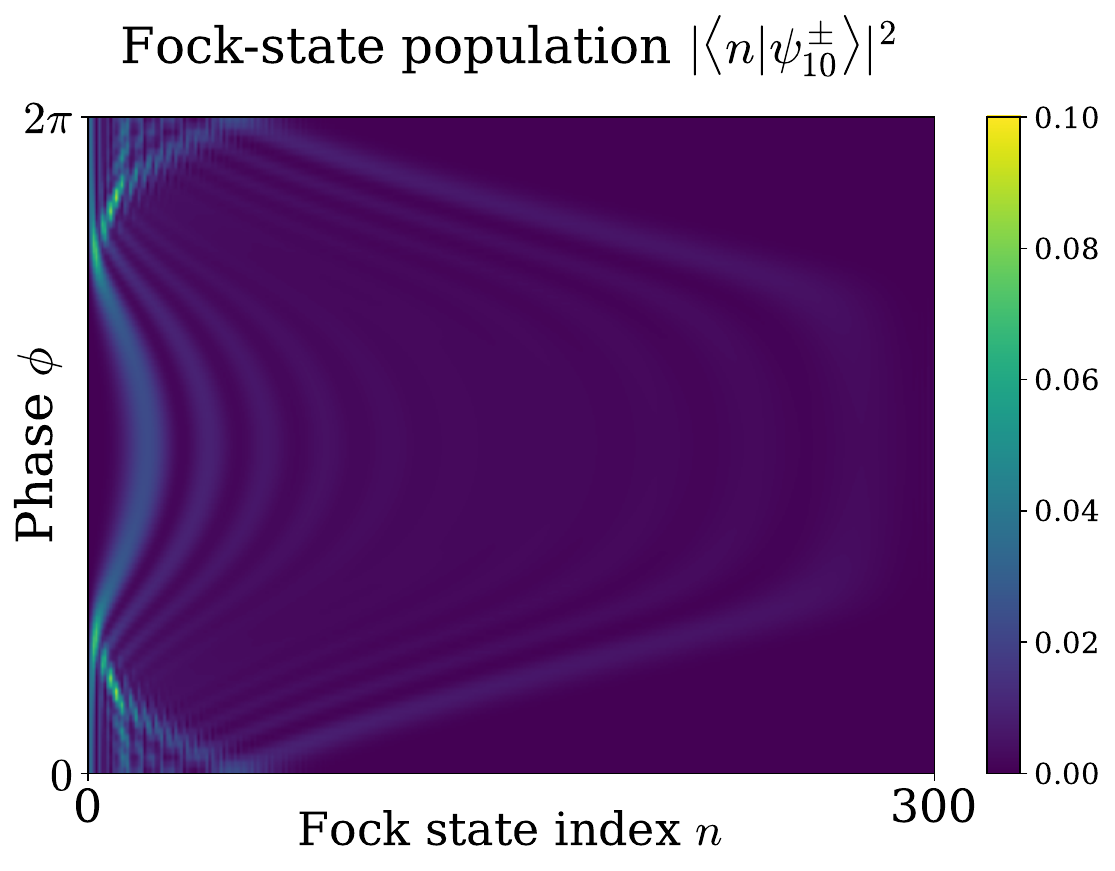}};
\node at (-2.2, 1.7) {\small (b)};
\end{tikzpicture}
\hfill
\begin{tikzpicture}
\node at (0,0) {\includegraphics[width=.3\columnwidth]{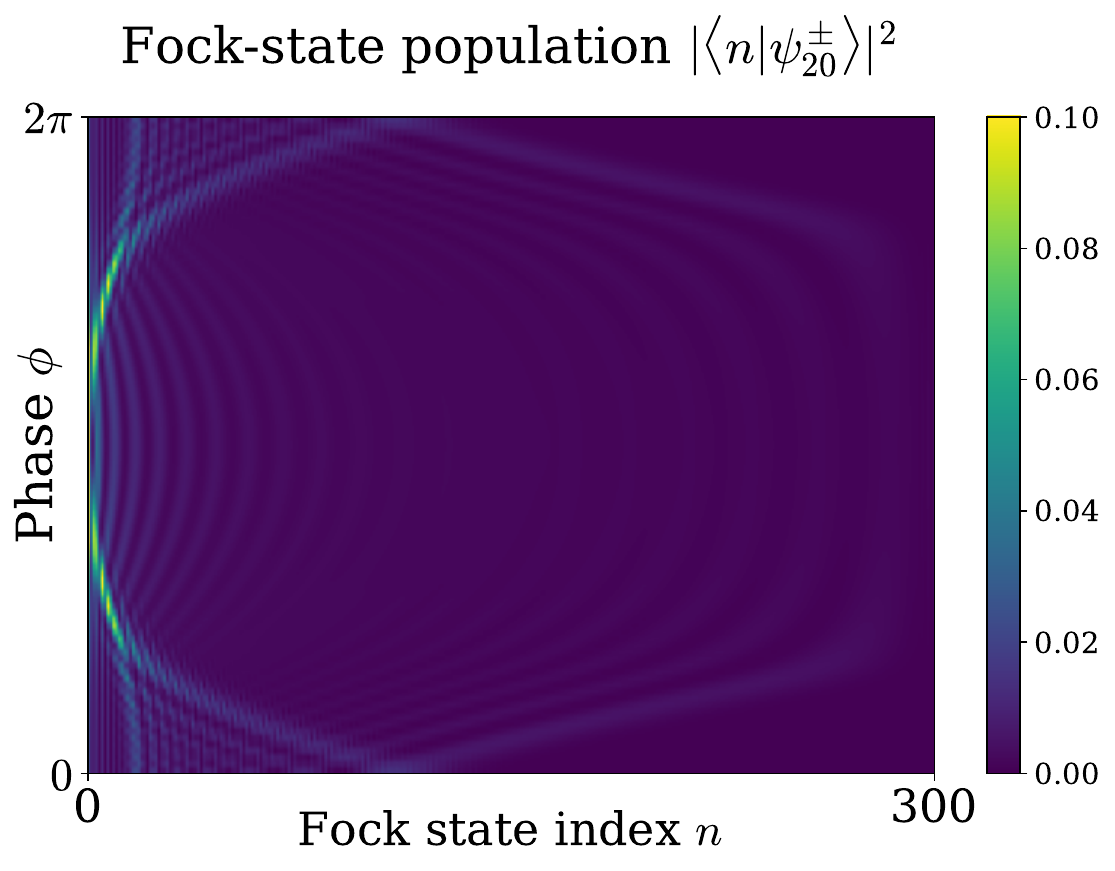}};
\node at (-2.2, 1.7) {\small (c)};
\end{tikzpicture}
\caption{
Fock-state distributions of (a) the zero-energy defect state, (b) \( |\psi_{10}^{\pm} \rangle \), and (c) \( |\psi_{20}^{\pm} \rangle \).  
The populations \( |\langle n | \psi \rangle|^2 \) at each Fock basis state are plotted for \( w = 2 \), \( \Delta v = 0.2 \), and a Fock-space cutoff of \( N = 300 \).  
For \( \phi \ne 0 \), the states spread over the Fock-state lattice due to \(\phi\)-dependent displacement and squeezing, reaching their maximal extent at \( \phi = \pi \).
}
\label{fig:figures3}
\end{figure}

The global marker is defined as the average of the local marker values within the bulk region, \( \mathbb{E}[\hat{C}(n)] \). 
As shown in Fig.~\ref{fig:figures2}(a), for \(\phi=0\) this quantity remains invariant over different \(w\) and exhibits a sharp transition at \(\Delta v=0\), independent of \(w\).
For \(\phi=\pi/2\), however, the global marker becomes sensitive to \(w\) since finite-size fluctuations near the edges affect the bulk average.
In this case, the marker deviates most strongly from its quantized value near the critical point \(|v_r|=|v_{cr}|\). 
The deviation increases with \(w\), as a larger \(w\) requires a larger cutoff \(N_{\mathrm{cut}}\) for the marker to faithfully capture the topological features of the system.
Figure~\ref{fig:figures2}(b) shows that the transition for \(\phi=\pi/2\) sharpens when the Fock-space dimension is increased to \(N=500\), although residual deviations remain due to fluctuations near \(n=0\).
Therefore, a reliable estimate requires the marker to be evaluated at sites \(n > N_{\mathrm{cut}}\).

The \(\phi\) dependence of the topological markers originates from the \(\phi\)-dependent distribution of eigenstates in the Fock basis.
For nonzero \(\phi\), the amount of squeezing is unchanged, but the state is displaced further in phase space, reaching a maximum at \(\phi=\pi\). 
This \(\phi\)-dependent phase-space motion broadens the Fock-basis distribution, making the local topological markers increasingly sensitive to \(\phi\). 
Figure~\ref{fig:figures3}(a) shows the Fock-space distribution of the defect state \(|\psi_0\rangle\) for \(w=2\) and \(\Delta v=0.2\). 
The eigenstates \(|\psi_n^{\pm}\rangle\) for \(n=10\) and \(20\) are shown in Figs.~\ref{fig:figures3}(b) and (c), respectively. 
As \(\phi\) varies from \(0\) to \(\pi\), these distributions expand with increasing displacement, requiring a larger Fock-space size \(N\) for the markers to yield reliable values.
We note that, while the real-space topological marker reflects these phase-space characteristics, it does not by itself provide such a direct physical interpretation.

Moreover, the marker approach presents additional experimental challenges compared to our proposed topological invariant—the phase-space winding number.
Determining an appropriate Fock-space cutoff requires extra calibration/measurements, and projective measurements onto positive- and negative-energy eigenstates (i.e., unoccupied and occupied bands) would require developing new experimental tools.
In contrast, the phase-space winding number can be accessed directly via well-established techniques for oscillator-state tomography, leveraging the distinctive properties of the defect state (see Sec.~\ref{section4}).
An experimental protocol for its measurement is provided in Sec.~\ref{section5}.

\section{Numerical Simulations}
\label{section4}
In this section, we present details regarding the simulation results plotted in Fig.~3 of the main text. 

\subsection{Adiabatic State Preparation}

In the simulation that prepares nonclassical bosonic states under noise (Fig.~3(a) in the main text), we consider an adiabatic process starting from the initial state \(|\psi_i\rangle = |g\rangle |0\rangle\). 
Initially, the parameters are set to \(|v_r|=1\), \(w=|v_{cr}|=0\), and \(\phi_r=\phi_{cr}=0\), and then we increase \(w \to 1\) and \(|v_{cr}| \to 0.25\). 
To make the evolution time \(T_{adia}\) satisfy the adiabatic condition when preparing the state \(|\psi_n^{\pm}\rangle\), we set \(T_{adia}\) sufficiently larger than the inverse of the minimal energy gap \(\Delta E_n\):
\begin{equation*}
\Delta E_n = \left|E_{n+1}^{\pm}- E_{n}^{\pm} \right|=  \sqrt{\left(|v_r|-|v_{cr}|\right)(n+1)} -      \sqrt{\left(|v_r|-|v_{cr}|\right)n}.
\end{equation*}
For example, when preparing \(|\psi_0\rangle\), the minimal energy gap is \(\Delta E_0 =E_{\pm1} - E_0 = \sqrt{|v_r|-|v_{cr}|}\), and we set \(T_{adia}=\frac{25}{\sqrt{|v_r|-|v_{cr}|}}\). 
For the other states \(|\psi_n^{\pm}\rangle\), we set \(T_{adia}\) in the same way.
We summarize the values and definitions of the parameters used for the simulation in Table.~\ref{tab:tableS1}

\begin{table}[h!]
\caption{System parameters used in simulation.}
\newcolumntype{D}{p{0.275\linewidth}}
\newcolumntype{N}{p{0.3\linewidth}}
\label{tab:tableS1}
\centering
\renewcommand{\arraystretch}{1.15}
\begin{tabularx}{0.98\linewidth}{c  D  c  c  N}
\toprule
\textbf{Symbol} & \textbf{Description} & \textbf{Value} & \textbf{Units} & \textbf{Notes} \\
\midrule
$N$        & Fock-space size                  &  100   & --  & Fixed \\
$w$        & Spin interaction                 &  1.0   & arb.& Adiabatically tuned from 0 \\
$|v_r|$    & Modulus of $v_r$                 &  1.0   & arb.& Fixed \\
$|v_{cr}|$ & Modulus of $v_{cr}$              &  0.25   & arb.& Adiabatically tuned from 0 \\
$\phi_r$   & Phase of $v_r$                  &  0     & rad & Fixed \\
$\phi_{cr}$& Phase of $v_{cr}$                  &  0     & rad & Fixed \\
$T_{adia}$ & Adiabatic evolution time         &  25/$\Delta E_n$    & arb. & Fixed \\
$\delta v$ & Noise amplitude  &  0-0.8    & arb. & Scanned over fixed range \\
$f_{\min}$ & Minimum noise frequency  &  0.1    & arb. & Fixed \\
$f_{\max}$ & Maximum noise frequency  &  2/$\Delta E_3$    & arb. & Fixed \\
\bottomrule
\end{tabularx}
\end{table}

\begin{figure}[t!]
\centerline{\includegraphics[width=0.75\columnwidth]{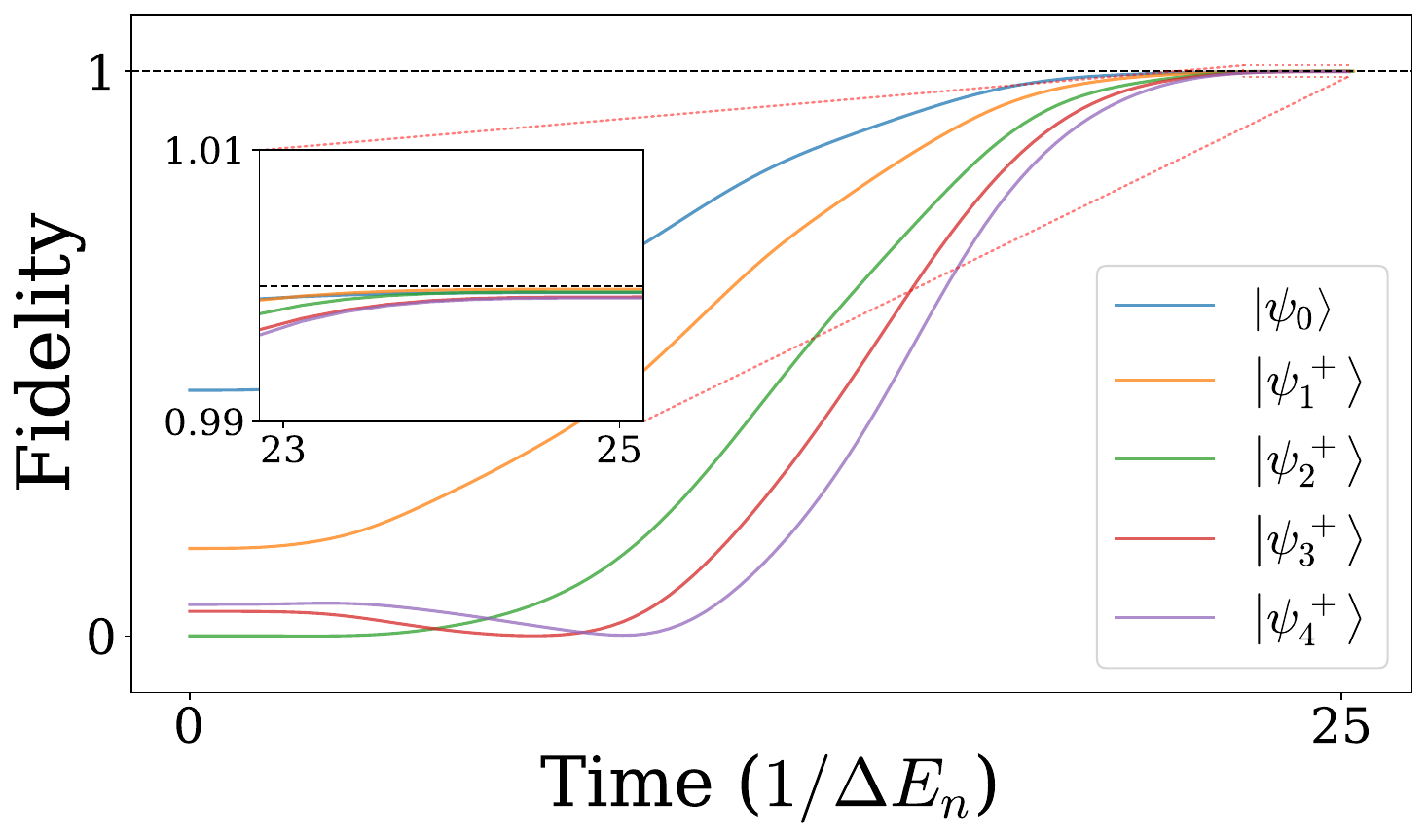}}
\caption{
Fidelity between the adiabatically prepared states and the target eigenstates \(|\psi_0\rangle\) and  \(|\psi_k^{+}\rangle\) (\(k \in \{1,2,3\}\)). 
The parameters \(w\) and \(v_{cr}\) are ramped according to the functions in Eq.~(\ref{eqn:eqn4})
}
\label{fig:figures4}
\end{figure}

In an ideal environment without the noise, we observe that the states \(|\psi_0\rangle\) and \(|\psi_k^{+}\rangle\) ($k\in\{1,2,3\}$) can be prepared with fidelity over 0.99, as shown in Fig.~\ref{fig:figures4}. 
The parameters are ramped according to the following functions:
\begin{align}
w(t) = \sin^2{\left(\frac{t}{T_{adia}}\right)}, \quad
v_{cr}(t) = 0.25\times\sin^2{\left(\frac{t}{T_{adia}}\right)}.
\label{eqn:eqn4}
\end{align}
The fidelities are computed at each time as:
\begin{align*}
\mathcal{F}(t) = \left| \langle \psi_n | \psi(t) \rangle \right|^2, \quad n\in\{0,1,2,3\}.
\end{align*}
We ignore \(+\) sign in the eigenstates for convenience.

\begin{figure}[b!]
\centerline{\includegraphics[width=0.8\columnwidth]{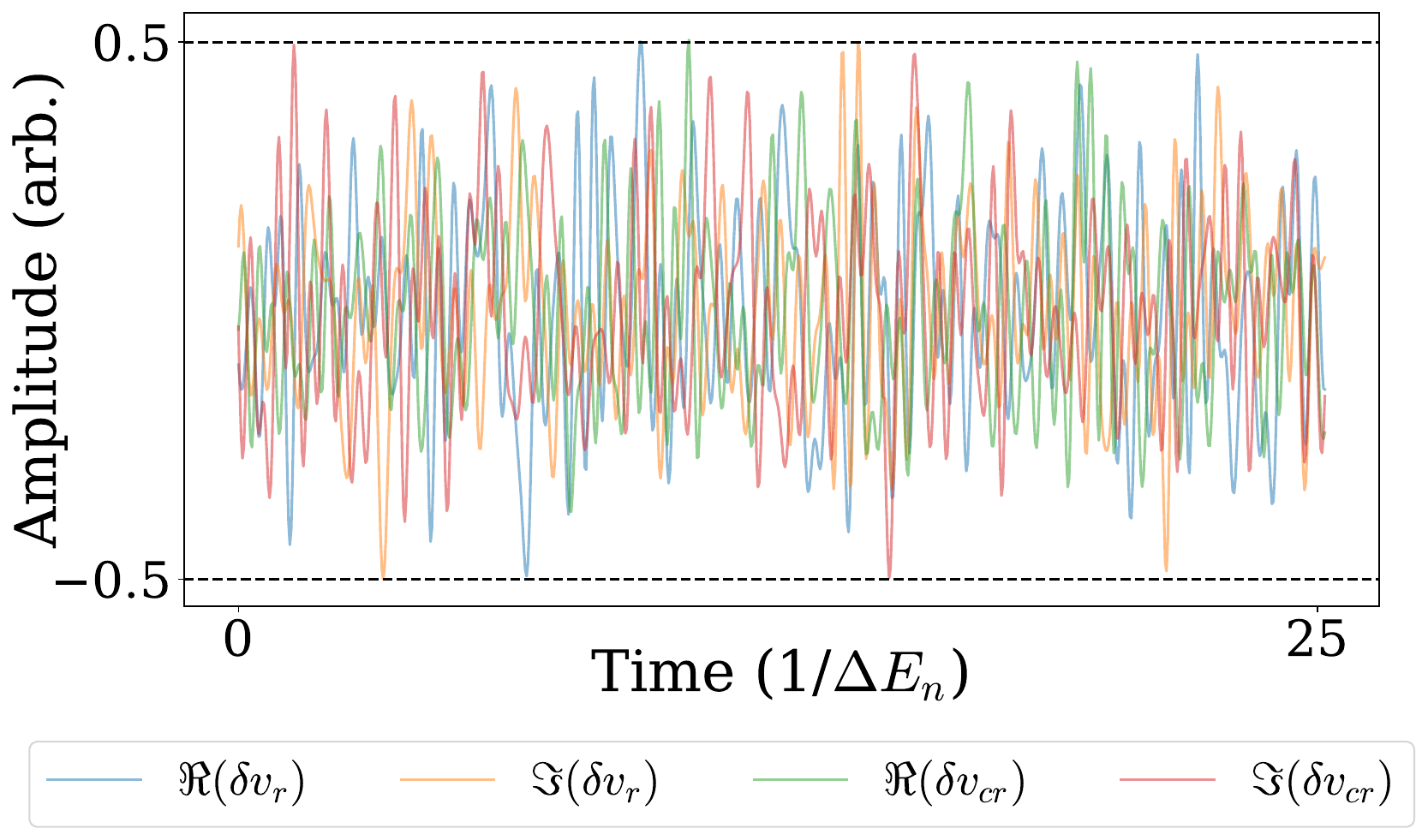}}
\caption{
Example noise profiles applied to \(v_r\) and \(v_{cr}\) during adiabatic preparation of the defect state \(\ket{\psi_0}\).
The amplitude is scaled by the parameter \(\delta v=0.5\).
The noise profiles are generated independently by sampling frequencies and phases as in Eq.~(\ref{eqn:eqn5}).
}
\label{fig:figures5}
\end{figure}

\begin{figure}[t!]
\centerline{\includegraphics[width=0.8\columnwidth]{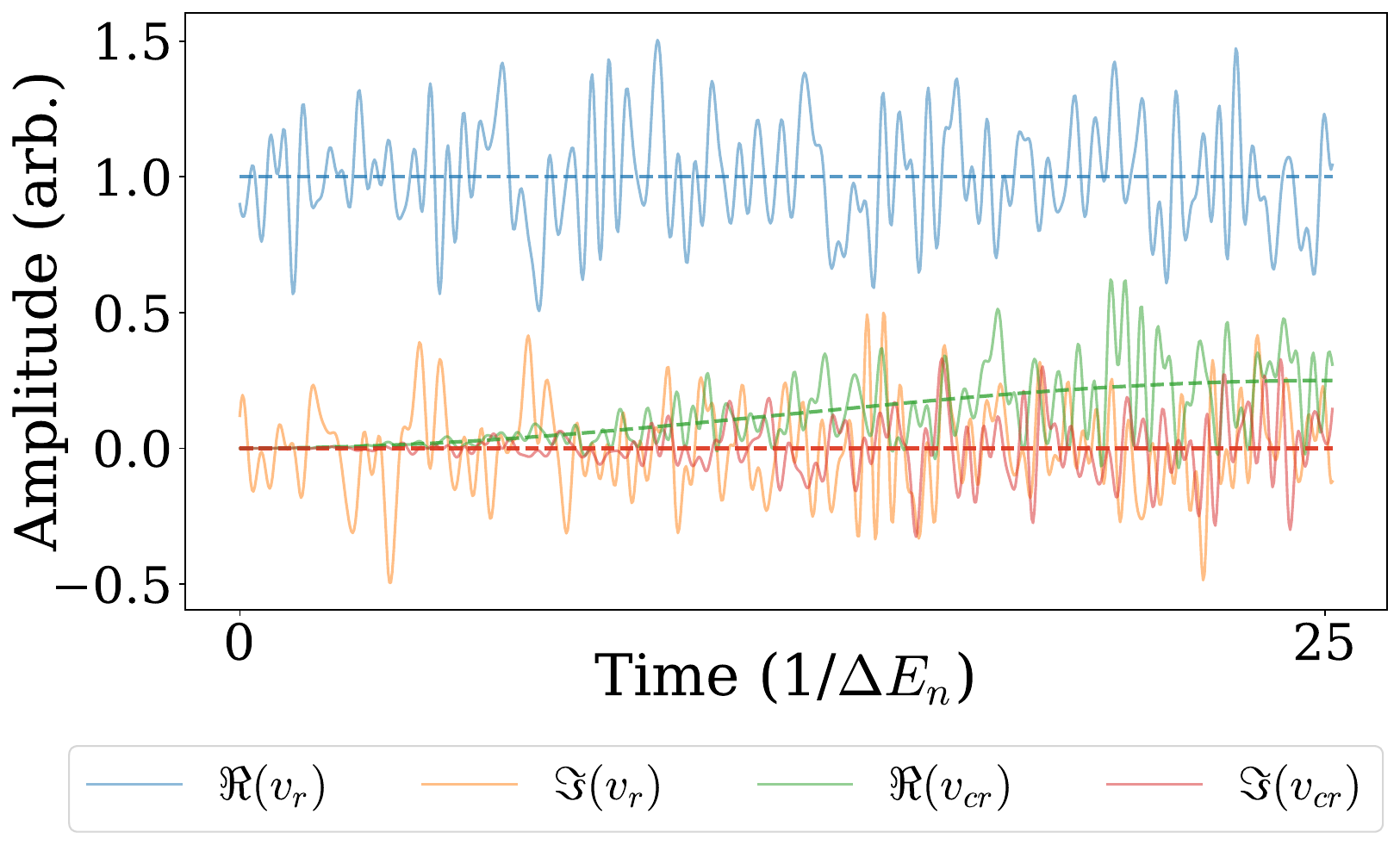}}
\caption{
Actual parameter trajectories under noise during the adiabatic process. Noise profiles from Fig.~\ref{fig:figures5} are added to \(v_r\) and \(v_{cr}\). \(v_{cr}\) is ramped according to Eq.~(\ref{eqn:eqn4}).
Dashed lines indicate the parameter values without noise.
}
\label{fig:figures6}
\end{figure}

\begin{figure}[b!]
\centerline{\includegraphics[width=0.75\columnwidth]{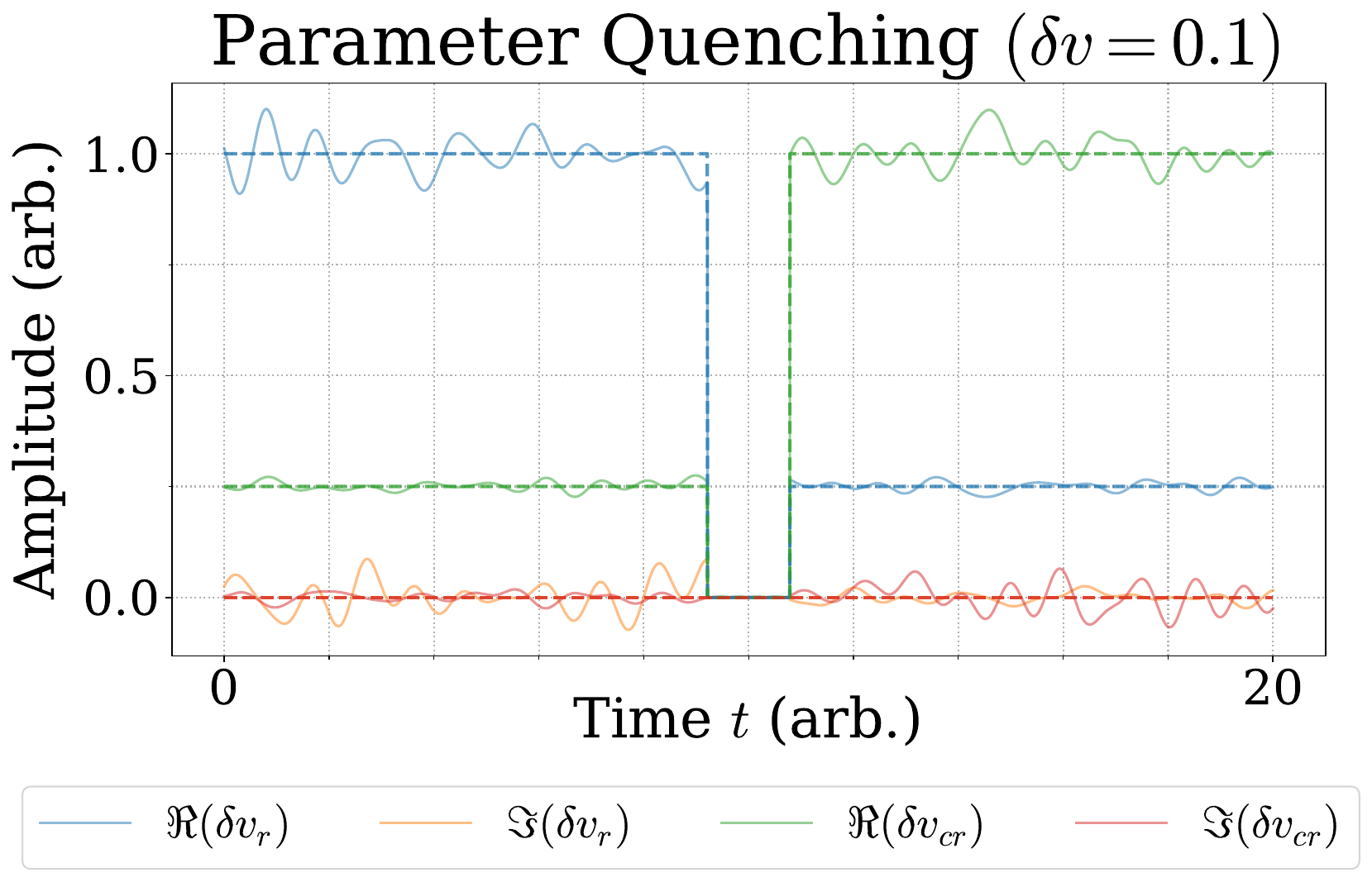}}
\caption{
Parameter trajectories used for the spin-flip operation.
Noise fluctuations acting on \(v_r\) and \(v_{cr}\) are generated using Eq.~(\ref{eqn:eqn5}) with amplitude \(\delta v=0.1\).
Dashed lines indicate the corresponding noiseless quenches.
}
\label{fig:figures7}
\end{figure}

The noise processes \(\delta v_r(t)\) and \(\delta v_{cr}(t)\) act additively on the corresponding parameters,
\begin{align*}
v_r(t) \to v_r(t) + \delta v_r(t), \qquad
v_{cr}(t) \to v_{cr}(t) + \delta v_{cr}(t).
\end{align*}
Each fluctuation is synthesized as a sum of sinusoids whose frequencies and phases are sampled uniformly,
\(f_i \sim \mathcal{U}(f_{\min},f_{\max})\) and \(\phi_i \sim \mathcal{U}(0,2\pi)\).
We set \(f_{\min}=0.1\) and \(f_{\max}=2/\Delta E_3\), since components much faster or much slower than the characteristic gap scale \(\Delta E_3\) have little influence on the adiabatic dynamics.

Noise acting on the real and imaginary parts of \(v_r\) and \(v_{cr}\) is generated independently and then normalized to a fixed amplitude window.
For example, for the real part of \(v_r\) we first generated a signal:
\begin{align}
\delta v_{r}(t) \;=\; \sum_{i=1}^{n_f} \sin\!\bigl(2\pi f_i t + \phi_i\bigr),
\qquad
f_i \sim \mathcal{U}(f_{\min},f_{\max}), \quad \phi_i \sim \mathcal{U}(0,2\pi),
\label{eqn:eqn5}
\end{align}
and then rescale it to have zero mean and peak amplitude \(\delta v\) over the simulation time grid \(\{t_m\}\):
\begin{align*}
\delta v_{r}(t_m) \;\leftarrow\;
\delta v \times
\frac{ \delta v_{r}^{(\mathrm{R})}(t_m) - \langle \delta v_{r}^{(\mathrm{R})}\rangle_{t} }
{ \max_{m}\!\bigl|\delta v_{r}^{(\mathrm{R})}(t_m) - \langle \delta v_{r}^{(\mathrm{R})}\rangle_{t}\bigr| },
\end{align*}
where \(\langle \cdot \rangle_{t}\) denotes the average over \(\{t_m\}\) and we use \(n_f=100\) in the simulations.
The same procedure is applied to the imaginary part of \(v_r\) and to both parts of \(v_{cr}\).
Example noise profiles \(\delta v_r(t)\) and \(\delta v_{cr}(t)\) used for preparing \(\ket{\psi_0}\) are shown in Fig.~\ref{fig:figures5}.
These noise signals are added to the parameters \(v_r\) and \(v_{cr}\), as shown in Fig.~\ref{fig:figures6}.
The dashed lines indicate the parameter values without noise.
For each Monte Carlo trial, the same noise profile was used with different noise amplitudes \(\delta v\).
The results in Fig.~3 of the main text show averages over 20 Monte Carlo trials with independently generated noise realizations.

For simulations including the Lindblad jump operator \(\hat{\sigma}_-\), we used QuTiP’s \texttt{mesolve} method~\cite{qutip}.
This solver simulates the Lindblad master equation
\begin{align*}
\dot{\rho}(t) = -i[H(t),\rho(t)] + \sum_n\frac{1}{2}\left[ 2 C_n \rho(t) C_n^{\dagger} - \rho(t) C_n^{\dagger} C_n - C_n^{\dagger} C_n \rho(t) \right], \quad C_n=\sqrt{\Gamma_n} A_n,
\end{align*}
where \(C_n\) are the jump operators.
In our numerics we set \(n=1\) and \(C_1=\sqrt{\Gamma}\,\hat{\sigma}_-\) with \(\Gamma=0.25\).

\subsection{Spin-Flip Operation}

\begin{figure}[b!]
\centerline{\includegraphics[width=0.8\columnwidth]{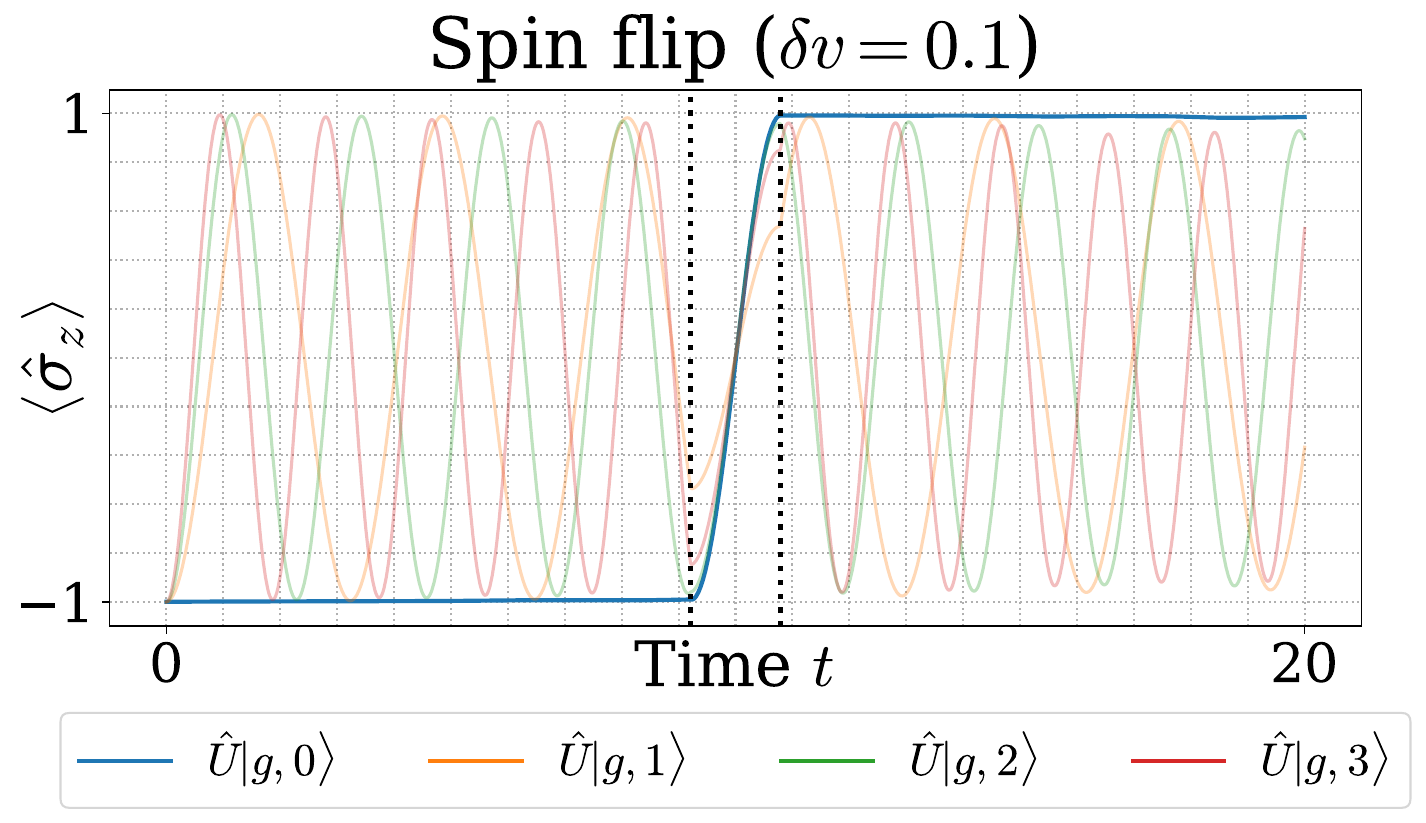}}
\caption{
Expectation value \(\langle \hat{\sigma}_z\rangle\) under the time-dependent parameters of Fig.~\ref{fig:figures7}.
Initial states are \(\hat{U}\ket{g,n}\) with \(n\in\{0,1,2,3\}\).
Only the defect state \(\hat{U}\ket{g,0}\) exhibits a robust, quantized spin-flip operation.
The other states display oscillations due to dynamical phases.
Vertical dotted lines indicate the quench times.
}
\label{fig:figures8}
\end{figure}

\begin{figure}[t!]
\centerline{\includegraphics[width=0.8\columnwidth]{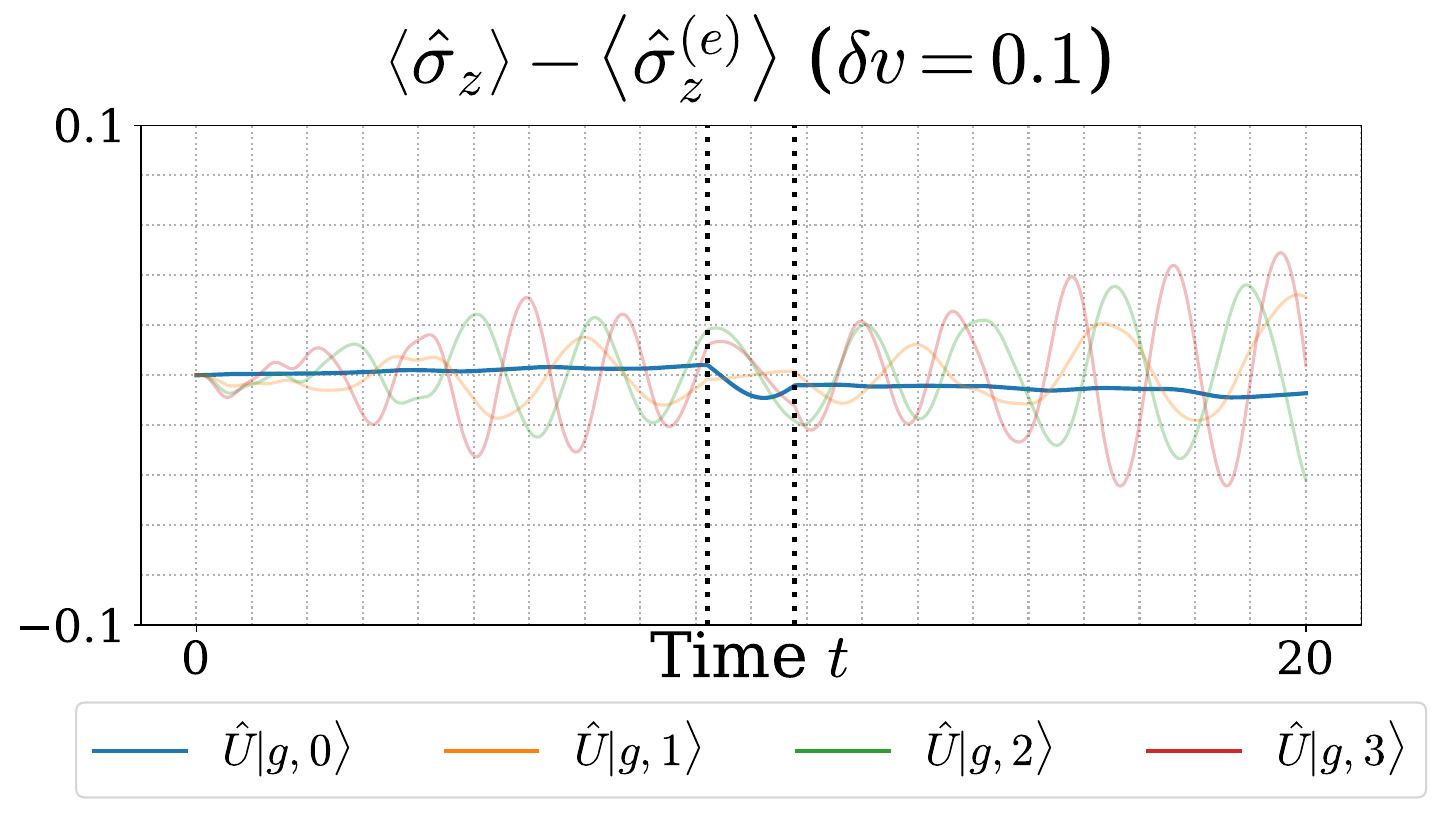}}
\caption{
Deviation of the spin polarization \(\left\langle\hat{\sigma}_z\right\rangle\) in Fig.~\ref{fig:figures8} from the noiseless expectation \(\langle\hat{\sigma}_z^{(e)}\rangle\), \(\left\langle\hat{\sigma}_z\right\rangle - \langle\hat{\sigma}_z^{(e)}\rangle\).
The defect state \(\hat{U}\ket{g,0}\) remains close to its noiseless value before and after the quench, whereas the states \(\hat{U}\ket{g,n}\) with \(n\neq 0\) exhibit larger deviations due to parameter noise.
}
\label{fig:figures9}
\end{figure}

In Fig.~3(b) of the main text, we demonstrate a spin-flip operation for the defect state \(\ket{\psi_0}\) and for the eigenstates \(\ket{\psi_k^{+}}\) (\(k\in\{1,2,3\}\)).
In the simulations, the system is initialized in the exact eigenstates \(\ket{\psi_0}\) or \(\ket{\psi_k^{+}}\), obtained by diagonalizing the Hamiltonian at \(v_r=w=1\) and \(v_{cr}=0.25\).
Noise profiles are generated as in the previous section using Eq.~(\ref{eqn:eqn5}) and are illustrated in Fig.~\ref{fig:figures7}.

We also simulate the states \(\hat{U}\ket{g,n}\) (\(n\in\{0,1,2,3\}\)) in the Fock-state lattice (FSL), with \(\hat{U}\) defined in the main text.
These states decompose into superpositions of the eigenstates \(\ket{\psi_k^{\pm}}\) and, in particular, \(n=0\) corresponds to the defect state \(\ket{\psi_0}=\hat{U}\ket{g,0}\).
Figure~\ref{fig:figures8} shows the evolution of the spin polarization \(\langle \hat{\sigma}_z \rangle\) under the time-dependent parameters of Fig.~\ref{fig:figures7}.
While \(\hat{U}\ket{g,0}\) exhibits the same dynamics as \(\ket{\psi_0}\) in the main text, the states \(\hat{U}\ket{g,n}\) with \(n\neq 0\) display oscillations due to dynamical phases accumulated by the contributing eigenstates \(\ket{\psi_k^{\pm}}\).
Therefore, only the defect state localized at the domain wall demonstrates a robust, quantized spin-flip operation.

The results in Fig.~\ref{fig:figures8} are compared with the corresponding noiseless spin polarizations \(\langle \hat{\sigma}_z^{(e)} \rangle\), as shown in Fig.~\ref{fig:figures9}.
The defect state \(\hat{U}\ket{g,0}\) remains robust before and after the quench, showing small deviations under noise.
In contrast, the states \(\hat{U}\ket{g,n}\) with \(n\neq 0\) in the FSL exhibit larger fluctuations in \(\langle \hat{\sigma}_z \rangle\) due to parameter noise.

To simulate the time-dependent dynamics in this section, we used the open-source Python library QuTiP~\cite{qutip}.
The noisy parameter trajectories shown in Fig.~\ref{fig:figures6} and \ref{fig:figures7} were interpolated with cubic splines to produce callable coefficient functions for the solver.
The Python scripts used for the simulations are available in the GitHub repository~\cite{git_repo}.

We note that such robust spin modulation can be used to encode information in the spin degree of freedom, while the bosonic mode in the system remains tunable.
Moreover, adiabatic preparation of the defect state \(\ket{\psi_0}\) together with the robust spin polarization \(\langle \hat{\sigma}_z\rangle\) facilitates experimental access to the topological invariant proposed in the main text---the phase-space winding number---via direct phase-space tomography based on parity measurements.
Details of the experimental protocol are provided in the following section.

\section{Experimental Measurements}
\label{section5}

\begin{figure}[b!]
\centerline{\includegraphics[width=1.\columnwidth]{figure_supplementary/FigureS10.pdf}}
\caption{
Experimental protocol for preparing and measuring the defect state.
(a) Adiabatic preparation of the defect state \(\ket{\psi_f}\) starting from \(\ket{\psi_i}=\ket{0}\ket{g}\).
The control parameters \(\phi\), \(\eta\), and \(\gamma\) are tuned experimentally (e.g., via the relative phase and amplitude of the sideband drives), thereby realizing defect-state parameters \((\theta,\alpha,r)\) within the accessible range.
After preparation, oscillator–state information is obtained via a conditional operation followed by spin readout.
Before measurement, the spin input is \(\ket{\psi_g}=\ket{g}\) and the bosonic state is \(\ket{\psi_b}=\hat{R}(\theta)\hat{D}(\alpha)\hat{S}(r)\ket{0}\).
(b) Measurement of the characteristic function via a spin-dependent displacement. 
The sign of the displacement depends on the spin states in the \(\hat{X}\) basis, denoted by the diamond symbol. 
See Ref.~\cite{tomography_ion} for more details.
(c) Measurement of the parity expectation via a conditional cavity phase shift; see Refs.~\cite{circuit_tomography,circuit_tomography2,matt_thesis}.
These measurements require a well-defined, non-oscillating spin during the preparation and readout sequence, a condition satisfied by the defect state.
}
\label{fig:figures10}
\end{figure}

In Section~\ref{section4} and Fig.~3 of the main text, we show that the defect state \(\ket{\psi_0}\) features a tunable bosonic mode and a quantized spin state, both robust to parameter noise.
These features enable the generation of nonclassical bosonic states for quantum metrology, while the spin degree of freedom robustly encodes digital information.
In this section, we further demonstrate that, by combining these properties, the topological invariant---the phase-space winding number---can be measured directly via oscillator-state tomography.

As shown in Section~\ref{section4}, the defect state can be prepared with high fidelity by adiabatically tuning the system parameters.
Without loss of generality, we consider the regime \(|v_r|>|v_{cr}|\) and fix \(v_r = |v| e^{i\phi}\).
Starting from \(\ket{\psi_i}=\ket{0}\ket{g}\), we ramp \(w: 0 \to \eta|v|\) and \(v_{cr}: 0 \to \gamma|v|\) to adiabatically prepare the defect state, as illustrated in Fig.~\ref{fig:figures10}(a).
By varying \(\phi\), \(\eta\), and \(\gamma\), the defect–state parameters \((\theta,\alpha,r)\) can be tuned to realize arbitrary target values within the accessible range.
The parameters \(\phi\), \(\eta\), and \(\gamma\) correspond to experimentally controllable settings---for example, the relative phase and amplitude of sidebands in a trapped-ion system~\cite{QRM_iontrap}.

After the defect state is prepared, the information of the oscillator state can be obtained through a conditional operation followed by a measurement on the spin state.
Among various techniques developed for oscillator-state tomography, we present two example protocols readily realizable in trapped-ion and superconducting-circuit platforms.

In trapped-ion systems, the spin-dependent force is a well-established method for generating spin–motion entanglement~\cite{ion_sdf}.
It is implemented by applying a bichromatic beam to a single ion, which realizes a spin-dependent displacement---the sign of the displacement depends on the spin state.
Recently, a simple protocol for measuring the characteristic function of a bosonic mode has been proposed, which utilizes an unconditional spin rotation followed by a spin-dependent displacement~\cite{tomography_ion}; see Fig.~\ref{fig:figures10}(b).
The characteristic function is obtained by measuring spin expectation values while sweeping the displacement parameter \(\beta\), 
The resulting function can be used to reconstruct the Wigner function of the bosonic state \(\ket{\psi_b}\) \cite{tomography_ion} and to locate the center of the defect mode in phase space.

In superconducting-circuit systems, the dispersive interaction is a standard tool for qubit readout~\cite{circuit_qed} and for measuring the parity of oscillator states~\cite{circuit_tomography,circuit_tomography2}.
As the parity expectation is directly related to the Wigner function~\cite{Wigner_parity}, the Wigner function can be reconstructed through projective parity measurements implemented with the dispersive Hamiltonian~\cite{matt_thesis}.
An example protocol for measuring the Wigner function (i.e., the parity expectation) is shown in Fig.~\ref{fig:figures10}(c).
The Wigner function is reconstructed by repeating the protocol while sweeping the displacement parameter \(\beta\).

By performing Wigner tomography---either via measurements of the characteristic function or of the parity expectation---over a range of \(\phi\), the phase-space winding number can be extracted by tracking the trajectory of the Wigner-function peak.

\bibliographystyle{supplemental}
\bibliography{supplemental}